\documentclass[%
 aip,
 jmp,%
 amsmath,amssymb,
 reprint,%
]{revtex4-1}

\usepackage{graphicx}
\usepackage{grffile}
\usepackage{dcolumn}
\usepackage{bm}
\usepackage{hyperref}
\usepackage{multirow}
\usepackage{color} 
\usepackage{etex}
\reserveinserts{58}
\usepackage{morefloats}

\usepackage{longtable}
\usepackage{float}
\usepackage{supertabular}

\usepackage{booktabs}
\usepackage{array}
 
\usepackage{listings}
\usepackage{xcolor}       
\definecolor{dkgreen}{rgb}{0,0.6,0}
\definecolor{gray}{rgb}{0.5,0.5,0.5}
\definecolor{light-gray}{rgb}{0.97,0.97,0.97}

\lstdefinelanguage{owlms}
    {morekeywords={xsd,owl,xml,dc,rdf,skos,description,PlainLiteral,int,float,
        some,only,value,min,exactly,max,and,or,not,
        Prefix,Ontology,Import,Individual,Facts,Types,Class,
        DataProperty,ObjectProperty,AnnotationProperty,Annotations,
        DifferentIndividuals,SubClassOf,EquivalentTo,DisjointWith,DisjointUnionOf,SubPropertyOf,DisjointClasses,DisjointProperties,
        Symmetric,Asymmetric,Reflexive,Irreflexive,Transitive,Functional,InverseFunctional,
        Characteristics,Range,Domain,Datatype},
     basicstyle=\scriptsize\ttfamily,
     backgroundcolor=\color{light-gray},
     keywordstyle=\color{blue},
     commentstyle=\color{gray},
     stringstyle=\color{dkgreen},
     numbers=left,
     numberstyle=\tiny\color{gray},
     stepnumber=1,
     numbersep=10pt,
     tabsize=2,
     showspaces=false,
     showstringspaces=false,
     breaklines=true,                           
     sensitive=true,                            
     morecomment=[l][commentstyle]{\#},         
     morestring=[b]",                           
}

\begin{document}

\preprint{XXXXX (preprint)}

\title[Textual differentiation in interaction networks]{A connective differentiation of textual production in interaction networks}

\author{Renato Fabbri}%
 \email{fabbri@usp.br}
  \affiliation{ 
Instituto de F\'isica de S\~ao Carlos, Universidade de S\~ao Paulo (IFSC/USP)
}%

\date{\today}

\begin{abstract}
This paper explores textual production in interaction networks, 
with special emphasis on its relation to topological measures. 
Four email lists were selected, in which measures were taken from the texts participants wrote. 
Peripheral, intermediary and hub sectors of these networks were observed to have discrepant linguistic elaborations.
For completeness of exposition, correlation of textual and topological measures were observed for the entire network
and for each connective sector. The formation of principal components is used for further insights of how measures are related.
\end{abstract}

\maketitle
\section{Introduction}
Textual production has received considerable attention from the social network analysis community.
Sentiment analysis and vocabularies related to different parties are among a number of examples~\cite{easley}.
The relation of topological and textual measures is the subject of this article, for the following reasons:
\begin{itemize}
    \item This relation has been set aside in literature, with scattered and vague suggestions of mutual implications of the text produced and topological characteristics of the agents in the network~\cite{easley}.
    \item This results eases understanding of human interaction, which is useful for both psychological and anthropological typologies (personality and cultural ``types'')~\cite{Deborah}.
    \item There are interesting hypothesis about verbal differentiation of network sections and groups, derived from a previous article by the same author~\cite{evoSN}, some of which were herein confirmed.
\end{itemize}

Next section exposes materials used for this research, its textual and network facets.
Section~\ref{sec:methods} explains the analysis roadmap, with the measures chosen and methods
for understanding data. Section~\ref{sec:results} is dedicated to detailing results and discussion.
Section~\ref{sec:remarks} has concluding remarks and further works envisioned. Appendices hold information about mailing lists, tables, results still to interpret and directions on data and scripts.

\section{Materials}\label{sec:materials}
Eighty thousand messages were analysed, twenty thousand from each email list (see Appendix~\ref{sec:si}).
This data was accessed online through the GMANE database~\cite{GMANE}. Each message
has an ID, the ID of the message it is a response to (if any), an author,
a ``date and time'' field registering the moment the message was sent, and the textual content.
Other fields are also available, but plays no direct role in the work here presented.
Basic information of messages, threads and authors are summarized in Table~\ref{tab:bas}.

\subsection{Network formation}\label{subsec:net}

\begin{figure}[h!]
    \centering
    \includegraphics[width=0.3\textwidth]{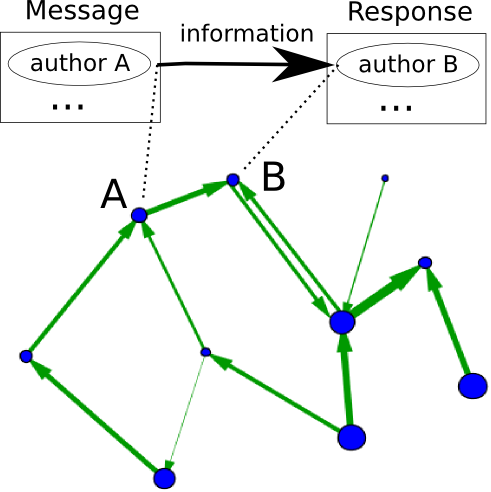}
    \caption{Formation of interaction network. Edges are directed as information flows, from
    an original message's author to the observed responder. Further information is given in Section~\ref{subsec:net}}
    \label{formationNetwork}
\end{figure}

Message-response pairs yield interaction networks, such as shown in Figure~\ref{formationNetwork}.
Each participant is represented by a vertex, and each response is considered evidence that
information emitted by the first author was received
by the responder (that had to read, process its contents and render a relevant textual response). Therefore, an edge from first author to the second author (responder) is considered.
This is the ``information network'' of the system. Edges can be considered in the reverse
order, from the responder to the original sender, representing status attribution, as the responder
considered what the sender said worthy of responding and is directing his attention to him. This is the ``status network''. As these networks
are virtually equivalent, one considers but one of them, usually the information network.

\subsection{Verbal observations}
Each message has a textual content. 
Analysis this content can consider author, network section or community, or messages independently.
As these are informal communities, there are typos, \emph{leet} and invented words.
This diversity and informality poses some challenges, by which the methodology was shaped.

\section{Methodology}\label{sec:methods}
An article was written for reporting stability in such networks from the topological viewpoint~\cite{evoSN}.
This article is dedicated to reporting differentiation in the textual production of the network as connectivity changes. Here, the observance of primary textual statistics is needed, and both overall incidences, and correlation to topological aspects, were tackled.

It is coherent to have participants as vertexes and as references for the messages sent, for the text produced and for activity (related to time and date). Thus, to observe the text produced in a certain section, one might gather all text produced by all participants on that section. To observe correlation of textual and topological characteristics, one can take measures on each vertex.

\subsection{Network measurements and partitioning}
Basic network measures of connectivity, in the same networks, were observed in a previous article~\cite{evoSN}.
The present article uses the same topological measures to observe correlations, PCA formation and network sectioning in peripheral, intermediary and hub sectors,
through strength measure. As described in that article, the ``exclusivist criteria'' for such partitioning is found to be the closest to literature
predictions (5\% of hubs, 15\% of intermediary and 80\% of peripheral vertex). Even so, strength-based
criteria is simpler and yields reasonable results (5-10\%, 5-25\%, 65-90\%).
Beyond that, changing the sectioning to a degree or a compound criteria did not significantly change the presented results.

Consequently, herein is considered a strength partitioning each sector (periphery, intermediary, hubs) is regarded as a \emph{primitive sector} of the \emph{primitive partitioning}.


\subsection{Textual measures}
An infinitude of textual measures can be drawn from texts. This work focuses on the simplest of them, as they proved sufficient for current step. These measures include frequency of individual letters and punctuations (Tables~\ref{tab:cha}), of words and tokens (Table~\ref{tab:tokens}), sizes of tokens, sentences and messages (Table~\ref{tab:sizesTokens},~\ref{tab:sizesSents} and~\ref{tab:sizesMsgs}) and POS (Part-Of-Speech) tags (Table~\ref{tab:pos}). Other measures envisioned are in subsection~\ref{subsec:fw}.

This choice is based on: 1) the lack of such information in literature, as far as authors know; 
2) potential relations of these incidences with topological aspects, such as connectivity; 3) the interdependence of textual artifacts suggests that simple measures should reflect complex behaviors subtle aspects.
A preliminary study, with all the work from Machado de Assis~\cite{letrasMachado}, made clear that these measures vary with respect to style.

Considered measures are:
\begin{itemize}
    \item Frequency of characters: letters, vowels, punctuations and uppercase. Table~\ref{tab:cha} is dedicated to such measures.
    \item Number of tokens, frequency of punctuations, of known words, of words that has wordnet synsets, of tokens that are stopwords, of words that return synsets and are stop words, etc. Table~\ref{tab:tokens} is dedicated to measures of this kind.
    \item Mean and standard deviation for word and token sizes. Table~\ref{tab:sizesTokens} is dedicated to these measures.
    \item Mean and standard deviation of sentence sizes. Table~\ref{tab:sizesSents} is dedicates to this sort of measures.
    \item Mean and standard deviation of message sizes. Table~\ref{tab:sizesMsgs} presents some of these measures.
    \item Fraction of morphosyntactic classes, such as adverbs, adjectives and nouns, represented by POS (Part-Of-Speech) tags. Table~\ref{tab:pos} displays such measures.
\end{itemize}

For sections (hubs, intermediary and peripheral),
all messages written by authors in each section were considered together. For the histograms, independent messages were considered from each sector. 

\subsection{Topological measures}\label{subsec:top}
Degree (in, out and total), strength (in, out and total), betweenness centrality and clustering coefficient were measured for each vertex in the interaction network. This served two purposes:
\begin{itemize}
    \item Obtaining a sound partitioning of the network in peripheral, intermediary and hub sectors. This was developed in a previous article by the same author~\cite{evoSN}.
    \item Observance of correlation with textual measures and principal components formation.
\end{itemize}

These measures are not developed here extensively as they are very consolidated, simple, and was the core of a previous article this subject by the same author~\cite{evoSN}.

\subsection{Relating text and topology}
Key observations for a deeper insight about network structure depend on
theoretical background and intentions. For this article, were considered:
\begin{enumerate}
    \item Incidences of linguistic traces in hub, intermediary and peripheral network sectors.
    \item Correlation of measures of each vertex, easing pattern detection involving topology of interaction and language used.
    \item PCA to gain further insights about how measures combine in principal components formation.
\end{enumerate}

Criteria for this choice include integration with previous topological results, lack of concise results in literature (as far as author knows) that
could substantiate correlations of topological and textual traces, and common sense as a long-time member of these networks.

First task, of textual production in hubs, intermediary and peripheral sectors, is observed by Tables~\ref{tab:cha}-\ref{tab:kolPctInter}. An adaptation of the Kolmogorov-Smirnov test was used to observe differences in textual productions, as exposed in Appendix~\ref{sec:ks}.

Second task is addressed by the correlation matrix with both textual and topological measurements of each participant, in Tables~\ref{tab:corTop}-\ref{tab:corTexTop}. Third, principal components composition are in Tables~\ref{tab:pca1}-\ref{tab:pca5}. 

\section{Results and discussion}\label{sec:results}

Although the results drawn from experiments and statistics were diverse, some fundamental insights can be obtained by going through tables and figures in the Appendix~\ref{sec:si}. Most importantly: connectivity has strong influence in textual production of participants in the network. For example: hubs use more contractions, more adjectives, more common words, and less punctuation if compared to the rest of the network, specially the peripheral sector. In general, rise or fall of a measure was monotonic, but some of them reached extreme values in the intermediary sector.

Next subsections exhibit particular results of interests.

\subsection{General characteristics of activity distribution among participants}\label{sec:gen}
Hubs and periphery swap fraction of participants and activity. While peripheral sector has $\approx 75\%$ of participants, it produces $\approx 10\%$ of all messages. Conversely, hubs has $\approx 10\%$ of participants and produces $\approx 75\%$ of all messages. Fewer threads are created in proportion to total messages sent by the hubs, while threads created by peripheral are twice as frequent as general peripheral messages.
This suggests a symbiosis of peripheral diversity and hub activity. 

Also, comparing lists with a fixed number of messages, the number of threads created seem to increase as the number of participants decrease. These information is condensed in Table~\ref{geralListas}, with further details.

\subsection{Characters}\label{sec:cha}
Peripheral vertex use more punctuation characters, digits and uppercase letters. Hubs use more letters and vowels among letters. The use of space does not seem to have any relation to connectivity, with the exception that the intermediary presented a slightly lower incidence of spaces than both peripheral and hub sectors. 
    
Total number of characters in ELE list, in the 20 thousand messages, is more than three times what other lists exhibited. This suggests peculiarities related to communication conventions and style (see Appendix~\ref{subsec:gm}).

Further information is given in Table~\ref{tab:cha}.

\subsection{Tokens and words}\label{subsec:tw}

Largest average size of tokens is with  the most wordy list (ELE). This implies that is has more characters, tokens, and characters per token in comparison to the other lists. Longer words used by hubs might be related to the use of a specialized vocabulary. Although the token diversity ($\frac{|tokens \neq|}{|tokens|}$) found in peripheral sector is far greater, this result has the masking artifact that the peripheral sector corpus is smaller, yielding a larger token diversity. This can be noticed by the token diversity of the whole network, which is lower than in the sections. This same discussion applies to the lexical diversity ($\frac{|kw\neq|}{kw}$).

Punctuations among tokens are less abundant in hubs, and discrepancies here are larger that with characters comparisons (subsection~\ref{sec:cha}). Known words are used more frequently by hubs.

ELE and CPP both exhibit intermediaries with the more frequent production of punctuation, less frequent production of known words, the highest incidence of words with wordnet synsets among known words. This suggests some peculiarity in network structure, such as the intermediary be strong authorities in such networks, using smaller sentences and a larger jargon.

Words with synsets, among known English words, are less frequent in hubs further evidencing the jargon hubs develop.

Further information is given in Table~\ref{tab:tokens}.

\subsection{Sizes of tokens and words}\label{subsec:tw2}
Sizes of known words are smaller for hubs, which suggests its use of more common words, although some of the previous results suggests that hubs have a very differentiated and specialized vocabulary. Larger words seems to be related to intermediary sector, which might be related to cultured vocabulary.

Further information is given in Table~\ref{tab:sizesTokens}.

\subsection{Sizes of sentences}\label{subsec:ss}
Hubs present the lowest average sentence size, both in characters and in tokens. Also, the incidence of usual known words seems to decay with connectivity, as does the number of known words with synsets.

Further information is given in Table~\ref{tab:sizesSents}.

\subsection{Messages}\label{subsec:mm}
Regarding characters and tokens, connectivity was related to smaller messages. ELE list displayed an inverse situation: the more connected the sector, the longer the messages are. This was considered a peculiarity of the culture bonded with the political subject of ELE list, to be further verified. Regarding sentences, the size of messages seem to hold steady until hubs are reached.

Further information is given in Table~\ref{tab:sizesMsgs}.

\subsection{POS tags}\label{subsec:pos}
Lower connectivity delivers more nouns and less adjectives, adverbs and verbs. This suggests that the networks collect issues important to the world by the peripheral sector. These issues are qualified, elaborated about, by the more connected participants.

Further information is given in Table~\ref{tab:pos}.

\subsection{Differentiation of measures}\label{subsec:di}
The Kolmogorov-Smirnov test was adapted for our need to compare measures.
Results suggests that the texts produced by each sector is very distinct.
Counterintuitively, intermediary sector sometimes yields a greater difference from periphery and hubs than these extreme sectors themselves (Tables~\ref{tab:kolSub} and~\ref{tab:kolSw}).

At the core of the results presented on this article, are two strong and immediate interpretations that follows Tables~\ref{tab:kolSub}-\ref{tab:kolPctInter}:
\begin{itemize}
    \item Differences of textual production of the primitive sectors are extreme, as can be noticed from the values on these tables, beyond reference values used for considering the null hypothesis (see Appendix~\ref{sec:ks}).
    \item Differences between sectors on the same network (Tables~\ref{tab:kolSub},~\ref{tab:kolAdj},~\ref{tab:kolSw} and~\ref{tab:kolPct}) are bigger than differences between same sector from distinct lists (Tables~\ref{tab:kolSubInter},~\ref{tab:kolAdjInter},~\ref{tab:kolSwInter} and~\ref{tab:kolPctInter}).
\end{itemize}

\subsection{Correlation of measures}\label{subsec:cor}

Correlation of degree (how many participants the participant related to) and strength (how many interactions) measures is substantially smaller for intermediary sector. This raises interesting inquiries, to which the reader is invited, along with further analysis of Tables~\ref{tab:corTop},~\ref{tab:corTex} and~\ref{tab:corTexTop}, as detailing their interpretation goes beyond the scope of this article. Noteworthy is the negative correlation of degree and message size (number of characters, tokens or sentences) that intermediaries presented.

\subsection{Formation of principal components}\label{subsec:pc}

Principal components formation seem to be the less stable of all features considered. First component, with $\approx 25\%$ of dispersion, relies heavily on POS tags, and slightly on sizes of tokens, sentences and messages. Second component, with almost $12\%$ of dispersion, blends topology, POS tags and size measures. Third component, with about $8.5\%$ is based on nouns frequency and size measures. Fourth and fifth components present less than $5\%$ of total dispersion, but are included for completeness of exposition.

Tables~\ref{tab:pca1}-\ref{tab:pca5} exhibit these and further insights.

\subsection{Results still to be interpreted}\label{subsec:sii}
These networks yield diverse characteristics, some of which were not of core importance for this step of the research. Even so, at least one of these characteristics was found interesting enough to be considered a result and an example of interesting artifacts found.

Histogram differences of incident and existent word sizes were found constant. That is, in each list, when a histogram of word sizes were made with all words written, and another histogram made with sizes of all \emph{different} words, the cumulative positive difference of the two histograms were found constant for all lists analysed. When all known English words were considered, the difference was always $\approx 1.0$. When stopwords were discarded, the difference found was different, but still constant, slightly above $0.5$. When only stopwords were considered, the difference was $\approx 0.6$. When only known English words that does not have wordnet synsets are used, this difference is $\approx 1.2$.

These results currently lacks substantial interpretation, which is provocative and should lead at least to a research note. Appendix ~\ref{sec:resE} and Figures~\ref{fig:kw}-\ref{fig:nssnsw} are dedicated to this histogram differences.

\section{Final remarks}\label{sec:remarks}
Human interaction networks yield diverse linguistic peculiarities reported by its members. This is a first systematic exploration of such peculiarities with primitive connective sectors (hubs, intermediary and periphery) in mind, as far the author knows. Results were regarded as stronger than envisioned from start, which poses diverse and intriguing questions.
This results, confluent with recent research and development, some by the current author, are of core importance for social technologies and transformations, such as collection and diffusion and information, resource recommendation in linked data contexts, and open processes of legal documents refinement~\cite{ensaio,OPS,pnud5,evoSN,pbr}.

All the data used is public, all scripts used are online (see Appendix~\ref{sec:on}). 

\subsection{Further work}\label{subsec:fw}
Results suggests that less connected participants bring external content and concepts, while hubs qualify the content. This hold mainly as periphery uses more nouns while hubs present more adjectives. This should be further verified, maybe with a dedicated article.

Similarity measures of texts in message-response threads has been thought about by the author, and some results are being organized. These are two hypothesis obtained from recent experiments:
\begin{itemize}
    \item existence of information ``ducts'', observable through similarity measures. These might coincide with asymmetries of edges between vertexes pairs, with homophily or with message-response threads, to point just a few possibilities.
    \item Valuable insights can be driven from self-similarity of messages by same authors, of messages sent at the same period of the day, etc. This includes incidences of word sizes, incidences of tags and morphosintactic classes, incidences of particular wordnet synset characteristics and wordnet word distances.
\end{itemize}

Given current results, diversity and self-similarity should vary with respect to connectivity. Literature usually assumes that periphery holds greater diversity~\cite{easley}, which should be further verified.

Other directions for next steps are:
\begin{itemize}
    \item Word sets are very useful to derive and confirm hypothesis. As an example, one can observe most incident or most basic words and word types in the English language, curses or words related to food.
    \item Interpretation of various unveiled results, such as the one exposed in Appendix~\ref{sec:resE}, and Figures~\ref{fig:kw},~\ref{fig:kwnsw},~\ref{fig:kwssnsw},~\ref{fig:sw} and~\ref{fig:nssnsw}.
    \item Extend word class observations to include plurals, gender, common prefixes and suffixes, etc.
    \item Date and time should also be addressed in textual production of interaction networks, as potentially linked to participation habits and purposes (e.g. low dispersion of sent time). This was tackled by the author for the topological characterization of interaction networks~\cite{evoSN}, but left aside in this article.
    \item Balance token diversity with corpus size, as pointed in section~\ref{subsec:tw}.
    \item The textual features distributions are likely to be have more than one peak or other non-trivial characteristic. Therefore, further analysis should be made for comparing measures of interest.
    \item Extend analysis to the windowed approach used in the article where hub, peripheral and intermediary sectors where topologically characterized~\cite{evoSN}.
    \item For ELE list, the more connected the sector, the longer the messages are. This is the inverse of what was found in the other lists, and was considered a peculiarity of the culture bonded with the political subject of ELE list, to be further verified.
    \item Tackle Portuguese analysis of interaction networks, as this research have ongoing implications in Brazil~\cite{ensaio}.
    \item Analyse other lists.
    \item Analyse interaction networks from other platforms such as Twitter, Facebook, LinkedIn, Diaspora, etc.
    \item Emotion classification has not been done and was considered out of the scope for this stage of development, but should be addressed in a near future.
\end{itemize}

Wordnet synsets incidences was studied as well, as a potentially useful way to characterize networks and sectors. Core aspects understood as useful for this research include:
\begin{itemize}
    \item Incidence of hypernyms, hyponyms, holonyms and meronyms.
    \item Use and development of similarity measures of words, phrases and messages, by use of semantic criteria (Wordnet) and bag of words.
\end{itemize}

\section{Acknowledgments}
Renato Fabbri is grateful to CNPq 
(processo 140860/2013-4, project 870336/1997-5);
Postgraduate Committee of the IFSC/USP;
Prof. Dr. Leonardo Paulo Maia for the insights
which lead to the adaptation of the Kolmogorov-Smirnov
test presented in Appendix~\ref{sec:ks}; GMANE
developers and maintainers; participants of the
mailing lists analysed.

%
\bibliographystyle{abbrv}
\bibliography{paper}  
%
%
\newpage
\appendix
\section{Adaptation of the Kolmogotov-Smirnov test}\label{sec:ks}
\subsection{Canonical Kolmogorov-Smirnov test}
Be $F_{1,n}$ and $F_{2,n'}$ two empirical distribution functions, where $n$ and $n'$ are the number of observations on each sample. The two-sample Kolmogorov-Smirnov test rejects the null hypothesis if:
\begin{equation}\label{eq:ks}
D_{n,n'} > c(\alpha)\sqrt{\frac{n+n'}{nn'}}
\end{equation}

where $D_{n,n'}=sup_x[F_{1,n}-F_{2,n'}]$ and $c(\alpha)$ is given for each level of $\alpha$:

\begin{table}[H]
\centering
\begin{tabular}{|l||c|c|c|c|c|c|}\hline
$\alpha$    & 0.1  & 0.05 & 0.025 & 0.01 & 0.005 & 0.001 \\\hline
$c(\alpha)$ & 1.22 & 1.36 & 1.48  & 1.63 & 1.73  & 1.95  \\\hline
\end{tabular}
\end{table}

\subsection{Adaptation}
We need to compare empirical distribution functions, so $D_{n,n'}$ is given, as are $n$ and $n'$. Therefore, as all terms in equation~\ref{eq:ks} are positive and $c(\alpha)$ can be isolated:

\begin{equation}\label{eq:ks2}
c(\alpha) < \frac{D_{n,n'}}{\sqrt{\frac{n+n'}{nn'}}} = c'(\alpha)
\end{equation}

Tables~\ref{tab:kolSub}-\ref{tab:kolPctInter} are populated with values for $c'(\alpha)$. When $c'(\alpha)$ is high, low values of $\alpha$ are possible for the test to reject the null hypothesis. Therefore, when $c'(\alpha)$ is greater than $\approx 1.7$, it is reasonable to assume that $F_{1,n}$ and $F_{2,n'}$ differ.

\section{Support information}\label{sec:si}
\subsection{Brief description of the email lists chosen}\label{subsec:gm}
GMANE is a public email list database with some tenths of thousand of lists~\cite{GMANE}. Four email lists were selected, in a similar fashion developed in~\cite{evoSN}, but with MET substituted by ELE list so that all lists are in English. The lists are:
\begin{itemize}
    \item CPP, the development list of the standard C++ library\footnote{gmane.comp.gcc.libstdc++.devel is list ID in GMANE archive.}. Dominated by specialized computer programmers.
    \item LAD: Linux Audio Developers list\footnote{gmane.linux.audio.devel is list ID in GMANE archive.}.
    \item LAU: Linux Audio Users list\footnote{gmane.linux.audio.users is list ID in GMANE archive.}.
    \item ELE: list for discussion of the election reform\footnote{gmane.politics.election-methods is list ID in GMANE archive}.
\end{itemize}

Table~\ref{tab:bas} has an overview of these lists, in terms of participants, threads and messages in each of the primitive connective sectors.

\subsection{Meaning of achronims and abbreviations in the following tables}

\begin{table}[H]
\centering
\begin{tabular}{|p{1cm}|p{6cm}|}\hline
symbol & meaning \\\hline\hline
$|x|$ & the number of times $x$ was found \\\hline
\emph{kw} & known word \\\hline
$|x\neq|$ & number of different $x$ found \\\hline
\emph{kwss} & known word with (wordnet) synset \\\hline
\emph{kwsw} & known word that is a stopword \\\hline
\emph{ukwsw} & unknown word that is a stopword \\\hline
\emph{nsssw} & word without (wordnet) synset that is a stopword \\\hline
\end{tabular}
\end{table}

Other symbols are explained on the tables itself. Some concepts, such as \emph{contractions}, \emph{token} and \emph{char} are standard in natural language processing, and the reader is invited to visit~\cite{nltkBook}.

\subsection{Tables}

\begin{table*}
\footnotesize
  \centering
\setlength{\tabcolsep}{.056667em}
  \begin{tabular}{|l|| c|c|c|c||  c|c|c|c||   c|c|c|c||   c|c|c|c|}\hline
-\-  & \multicolumn{4}{c|}{CPP} & \multicolumn{4}{c|}{LAD} & \multicolumn{4}{c|}{LAU} & \multicolumn{4}{c|}{ELE} \\ \hline
 & g. & p. & i. & h. &     g. & p. & i. & h. &    g. & p. & i. & h. &    g. & p. & i. & h. \\\hline
$date_1$                     & 3/13/2 &  - &  - &  - &  6/30/3 &  - &  - &  - & 06/29/3 &  - &  - &  - & 3/18/02 &  - &  - &  - \\
$date_M$                     & 8/25/9 &  - &  - &  - & 10/07/9 &  - &  - &  - & 07/23/5 &  - &  - &  - & 8/31/11 &  - &  - &  -    \\\hline
$N$  & 1052 & 834 & 163 & 55 & 1268 & 936 & 210 & 122 & 1183 & 904 & 155 & 124 & 302 & 225 & 36 & 41 \\
$N_{\%}$  & - & 79.28\% & 15.49\% & 5.23\% & - & 73.82\% & 16.56\% & 9.62\% & - & 76.42\% & 13.10\% & 10.48\% & - & 74.50\% & 11.92\% & 13.58\% \\\hline
M  & 19993 & 1654 & 2673 & 15666 & 19996 & 2331 & 3542 & 14123 & 19995 & 3018 & 2882 & 14095 & 19946 & 1821 & 2413 & 15712 \\
$M_{\%}$  & - & 8.27\% & 13.37\% & {\bf 78.33\%} & - & 11.65\% & 17.71\% & {\bf 70.61\%} & - & 15.09\% & 14.41\% & {\bf 70.47\%} & - & 9.11\% & 12.06\% & {\bf 78.56\%} \\\hline
$\Gamma$  & 4506 & 924 & 702 & 2880 & 3113 & 812 & 670 & 1631 & 3373 & 1121 & 675 & 1577 & 6070 & 782 & 1072 & 4216 \\
$\Gamma_{\%}$  & - & 20.51\% & 15.58\% & {\bf 63.91\%} & - & 26.08\% & 21.52\% & {\bf 52.39\%} & - & 33.23\% & 20.01\% & {\bf 46.75\%} & - & 12.88\% & 17.66\% & {\bf 69.46\%} \\\hline\hline
$-M$               & 7     &-&-&-      & 4 &-&-&-          & 5 &-&-&-           & 54 &-&-&-   \\ \hline
$\Delta_Y$ & 7.44 & - & - & - &                                  6.25  & - & - & - & 2.08  & - & - & - &     9.37 & - & - & - \\ \hline 
  \end{tabular}
  \caption{Columns $date_1$ and $date_M$ have dates (month/day/year) of first and last messages from the 20,000 messages considered. $N$ is the number of participants (number of different email addresses). $M$ is number of messages. $\Gamma$ is the number of threads (count of messages without antecedent). $-M$ is messages missing in the 20,000 collection, $100\frac{54}{20000}=0.27/100$ in the worst case. ELE notably has the fewer participants and the larger number of threads. This relation holds for pairs of lists considered: as the number of participants increase, the number of threads decrease. A similar role is observed in MET list described in~\cite{evoSN}, suggesting that 1) Non-technical topics gathers fewer participants and yields shorter threads; 2) MET techno-political characteristic is confirmed by having intermediary $\frac{N}{\Gamma}$ relation, between ELE (politics) and LAD (highly technical - GNU/Linux and music). These results should be further investigated in future research (see section~\ref{subsec:fw}). The number of threads started by hubs is significantly lower than activity for all list, this suggests creative exploitation is done by hubs, i.e. hubs acquire/absorb creativity. $\Delta_Y$ is number of years involved in the first 20,000 messages of each list. Dates of first and last message is: Mar/13/2002 and Aug/25/2009 for CPP; Jun/30/2003 and Oct/07/2009 for LAD; Jun/29/2003 and Jul/23/2005 for LAU; finally, Abr/18/2002 and Aug/31/2011 for ELE. See section~\ref{sec:results} and subsection~\ref{sec:gen} for further directions.}
  \label{geralListas}\label{tab:bas}
\end{table*}

\renewcommand{\arraystretch}{2}
\begin{table*}
  \centering
    \small
\setlength{\tabcolsep}{.056667em}
  \begin{tabular}{|l|| c|c|c|c||  c|c|c|c||   c|c|c|c||   c|c|c|c|}\hline
-\-  & \multicolumn{4}{c|}{CPP} & \multicolumn{4}{c|}{LAD} & \multicolumn{4}{c|}{LAU} & \multicolumn{4}{c|}{ELE} \\ \hline
 & g. & p. & i. & h. &     g. & p. & i. & h. &    g. & p. & i. & h. &    g. & p. & i. & h. \\\hline
$n\,chars$  & 12708286 & 11.65 & 17.65 & 70.69 & 12632264 & 14.21 & 18.21 & 67.58 & 11893325 & 17.37 & 15.60 & 67.04 & 38719505 & 7.74 & 11.17 & 81.09 \\\hline
$100\frac{|space|}{|char|}$  & 17.03 & 17.66 & 15.68 & 17.26 & 18.35 & 18.50 & 18.16 & 18.38 & 19.17 & 20.14 & 19.18 & 18.91 & 18.19 & 17.86 & 17.82 & 18.28 \\\hline
$100\frac{|punct|}{|char|-|space|}$  & 10.10 & 10.88 & 12.11 & 9.45 & 5.67 & 6.27 & 5.81 & 5.50 & 5.88 & 6.66 & 5.86 & 5.69 & 4.68 & 4.97 & 5.06 & 4.60 \\\hline
$100\frac{|digit|}{|char|-|space|}$  & 2.44 & 3.18 & 3.07 & 2.15 & 1.63 & 2.79 & 1.57 & 1.40 & 2.25 & 3.26 & 2.54 & 1.92 & 0.99 & 1.21 & 1.66 & 0.88 \\\hline
$100\frac{|letter|}{|char|-|space|}$  & 87.28 & 85.77 & 84.47 & 88.24 & 92.65 & 90.86 & 92.55 & 93.05 & 91.82 & 90.02 & 91.52 & 92.35 & 94.28 & 93.79 & 93.18 & 94.48 \\\hline
$100\frac{|vogal|}{|letter|}$  & 35.36 & 36.42 & 36.08 & 37.51 & 34.20 & 35.93 & 35.56 & 37.55 & 34.65 & 36.29 & 35.94 & 37.34 & 35.71 & 36.56 & 36.24 & 37.52 \\\hline
$100\frac{|Uppercase|}{|letter|}$  & 4.60 & 4.96 & 5.38 & 3.55 & 6.06 & 6.05 & 6.19 & 3.77 & 5.31 & 4.88 & 5.78 & 4.15 & 4.20 & 4.75 & 5.09 & 3.44 \\\hline
  \end{tabular}
  \caption{Measures based on characters of the text produced by network participants, fairly stable. Suggested relations are: 1) punctuations of CPP, that can be expected by its programming language focus and dots and semicolon abundance in such parlance; 2) greater number of letters on ELE is expected by its political disposition; 3) not statistically clear, but higher percentage of vowels might be a sign of erudition. Most of all, number of characters incident in ELE 20,000 messages are more then the other three lists added. MET has an intermediary value of 13,137,042 characters~\cite{evoSN}, above CPP, LAD, LAU and below ELE. This builds up to a dichotomic typology of networks: technical (more participants, fewer and longer threads, e.g. CPP) -- political (less participants, more and shorter threads, e.g. ELE). Higher incidence of digits and lower incidence of letters seem to be associated to technical subjects. See subsection~\ref{sec:cha} for further discussion and context.}
  \label{tab:cha}
\end{table*}

\begin{table*}
  \centering
    \footnotesize
\setlength{\tabcolsep}{.056667em}
  \begin{tabular}{|l|| c|c|c|c||  c|c|c|c||   c|c|c|c||   c|c|c|c|}\hline
-\-  & \multicolumn{4}{c|}{CPP} & \multicolumn{4}{c|}{LAD} & \multicolumn{4}{c|}{LAU} & \multicolumn{4}{c|}{ELE} \\ \hline
 & g. & p. & i. & h. &     g. & p. & i. & h. &    g. & p. & i. & h. &    g. & p. & i. & h. \\\hline
$|tokens|$ & 2839679  & 0.12 & 0.18 & 0.70 & 2686539  & 0.14 & 0.18 & 0.68 & 2588673  & 0.17 & 0.16 & 0.67 & 8019188  & 0.08 & 0.11 & 0.81 \\\hline
$\frac{|chars|-|spaces|}{|tokens|}$ & 3.71 & 3.63 & 3.69 & 3.73 & 3.84 & 3.83 & 3.84 & 3.84 & 3.71 & 3.68 & 3.70 & 3.72 & 3.95 & 3.93 & 3.89 & 3.96 \\\hline
$100\frac{|tokens \neq|}{|tokens|}$ & 1.84 & 5.60 & 4.08 & 1.89 & 2.43 & 6.55 & 5.10 & 2.67 & 2.42 & 5.64 & 5.40 & 2.73 & 0.85 & 3.45 & 2.83 & 0.86 \\\hline
$100\frac{|punct|}{|tokens|}$ & 26.48 & 27.59 & 29.96 & 25.39 & 17.96 & 19.89 & 18.37 & 17.45 & 18.29 & 20.57 & 18.58 & 17.63 & 16.35 & 17.07 & 17.32 & 16.14 \\\hline
$100\frac{|known\,words=kw|}{|tokens| - |punct|}$ & 76.52 & 71.24 & 70.27 & 78.89 & 84.42 & 80.43 & 83.47 & 85.49 & 81.92 & 76.30 & 80.69 & 83.60 & 90.01 & 88.58 & 86.72 & 90.60 \\\hline
$100\frac{|kw\neq|}{kw}$ & 0.83 & 3.83 & 2.81 & 1.00 & 1.06 & 3.96 & 3.16 & 1.36 & 1.11 & 3.36 & 3.50 & 1.44 & 0.43 & 2.55 & 1.97 & 0.49 \\\hline
$100\frac{|kw\,with\,wordnet\,synset=kwss|}{|kw|}$ & 74.21 & 74.75 & 75.74 & 73.80 & 74.84 & 75.26 & 74.80 & 74.76 & 74.80 & 75.54 & 75.02 & 74.59 & 73.49 & 73.77 & 74.16 & 73.38 \\\hline
100$\frac{|kw\,that\,are\,stopwords=kwsw|}{|kw|}$ & 47.14 & 46.02 & 44.32 & 47.91 & 49.16 & 46.62 & 48.64 & 49.78 & 49.26 & 46.86 & 48.44 & 49.98 & 49.25 & 48.43 & 48.16 & 49.47 \\\hline
$100\frac{|unknown\,words\,that\,are\,sw=ukwsw|}{|kw|}$ & 2.86 & 3.39 & 2.73 & 2.81 & 2.56 & 2.82 & 2.74 & 2.46 & 3.67 & 4.04 & 3.68 & 3.58 & 1.73 & 1.90 & 2.04 & 1.67 \\\hline
$100\frac{|kw\,that\, are\, stopwords\,and\,have\,synsets|}{|kw|}$ & 24.29 & 23.84 & 23.31 & 24.57 & 26.39 & 24.38 & 25.83 & 26.93 & 26.60 & 25.20 & 26.05 & 27.04 & 25.22 & 24.78 & 24.69 & 25.33 \\\hline
$100\frac{|stopwords\,without\,synsets|}{|kw|}$ & 22.85 & 22.18 & 21.01 & 23.34 & 22.76 & 22.24 & 22.81 & 22.85 & 22.66 & 21.67 & 22.39 & 22.94 & 24.03 & 23.65 & 23.47 & 24.14 \\\hline
$100\frac{|contractions|}{|kw|}$ & 1.65 & 1.24 & 1.59 & 1.72 & 1.76 & 1.34 & 1.59 & 1.89 & 2.19 & 1.73 & 1.74 & 2.40 & 1.43 & 1.26 & 1.33 & 1.46 \\\hline
$100\frac{|kw\,not\,stopwords\,no\,synset|}{|kw|}$ & 2.94 & 3.07 & 3.26 & 2.86 & 2.40 & 2.50 & 2.39 & 2.39 & 2.54 & 2.79 & 2.59 & 2.47 & 2.48 & 2.58 & 2.37 & 2.48 \\\hline
$100\frac{|kw\,not\,stopword\,has\,synset|}{|kw|}$ & 49.92 & 50.92 & 52.42 & 49.23 & 48.44 & 50.88 & 48.97 & 47.84 & 48.20 & 50.35 & 48.97 & 47.55 & 48.27 & 48.99 & 49.47 & 48.05 \\\hline
  \end{tabular}
  \caption{Basic measures on tokens, known English words, stopwords, words with and without synset. Lexical diversity is higher in LAU and LAD, probably linked to these lists hybrid technical interests (music and GNU/Linux). Larger known words and tokens are incident in ELE and LAD. ELE also exhibits larger incidence of stopwords without synsets (prolixity?). Stronger use of words with synsets that are not stopwords is held by CPP. Stopwords that have synset account for $\approx25\%$ of all known words, which might be an indicative of language complexity (not same as good writing though). See subsection~\ref{sybsec:tw} for further discussion and context.}
  \label{tab:tokens}
\end{table*}

\begin{table*}
\footnotesize
\setlength{\tabcolsep}{.56667em}
  \centering
  \begin{tabular}{|l|| c|c|c|c||  c|c|c|c||   c|c|c|c||   c|c|c|c|}\hline
-\-  & \multicolumn{4}{c|}{CPP} & \multicolumn{4}{c|}{LAD} & \multicolumn{4}{c|}{LAU} & \multicolumn{4}{c|}{ELE} \\ \hline
 & g. & p. & i. & h. &     g. & p. & i. & h. &    g. & p. & i. & h. &    g. & p. & i. & h. \\\hline
$\mu(size\,of\,known\,word=skw)$ & 4.51 & 4.53 & 4.56 & 4.50 & 4.44 & 4.52 & 4.45 & 4.42 & 4.35 & 4.42 & 4.36 & 4.34 & 4.64 & 4.65 & 4.66 & 4.63 \\
$\sigma(skw)$ & 2.39 & 2.38 & 2.42 & 2.39 & 2.35 & 2.40 & 2.36 & 2.34 & 2.25 & 2.27 & 2.25 & 2.25 & 2.52 & 2.54 & 2.53 & 2.51 \\
$\mu(\neq skw)$ & 7.52 & 7.15 & 7.29 & 7.50 & 7.54 & 7.24 & 7.22 & 7.51 & 7.43 & 7.02 & 7.09 & 7.41 & 7.92 & 7.62 & 7.69 & 7.91 \\
$\sigma(\neq skw)$ & 2.57 & 2.51 & 2.56 & 2.57 & 2.53 & 2.54 & 2.53 & 2.53 & 2.51 & 2.49 & 2.48 & 2.51 & 2.62 & 2.62 & 2.63 & 2.61 \\ \hline
$\mu(skwss)$ & 4.92 & 4.94 & 4.95 & 4.95 & 4.82 & 4.94 & 4.84 & 4.84 & 4.70 & 4.77 & 4.71 & 4.71 & 5.11 & 5.14 & 5.14 & 5.14 \\
$\sigma(skwss)$ & 2.54 & 2.52 & 2.56 & 2.56 & 2.50 & 2.54 & 2.50 & 2.50 & 2.40 & 2.40 & 2.38 & 2.38 & 2.69 & 2.70 & 2.68 & 2.68 \\
$\mu(\neq skwss)$ & 7.56 & 7.20 & 7.34 & 7.34 & 7.57 & 7.29 & 7.27 & 7.27 & 7.47 & 7.09 & 7.14 & 7.14 & 7.94 & 7.66 & 7.73 & 7.73 \\
$\sigma(\neq skwss)$ & 2.54 & 2.48 & 2.52 & 2.52 & 2.49 & 2.51 & 2.50 & 2.50 & 2.48 & 2.46 & 2.45 & 2.45 & 2.58 & 2.59 & 2.60 & 2.60 \\ \hline
$\mu(ssw)$ & 2.89 & 2.87 & 2.87 & 2.89 & 2.85 & 2.83 & 2.85 & 2.86 & 2.86 & 2.86 & 2.85 & 2.87 & 2.88 & 2.86 & 2.87 & 2.88 \\
$\sigma(ssw)$ & 1.06 & 1.06 & 1.07 & 1.06 & 1.06 & 1.05 & 1.05 & 1.06 & 1.05 & 1.05 & 1.04 & 1.05 & 1.09 & 1.09 & 1.09 & 1.09 \\
$\mu(\neq ssw)$ & 3.92 & 3.88 & 3.90 & 3.89 & 3.97 & 3.92 & 3.90 & 3.97 & 3.97 & 3.92 & 3.92 & 3.97 & 3.97 & 3.97 & 3.97 & 3.97 \\
$\sigma(\neq ssw)$ & 1.60 & 1.58 & 1.60 & 1.58 & 1.68 & 1.65 & 1.60 & 1.69 & 1.68 & 1.61 & 1.60 & 1.69 & 1.68 & 1.69 & 1.68 & 1.68 \\ \hline
$\mu(snsssw)$ & 3.01 & 2.98 & 2.99 & 3.02 & 2.97 & 2.96 & 2.96 & 2.98 & 2.99 & 2.99 & 2.97 & 2.99 & 2.99 & 2.97 & 2.97 & 2.99 \\
$\sigma(snsssw)$ & 1.25 & 1.23 & 1.25 & 1.26 & 1.25 & 1.24 & 1.23 & 1.25 & 1.25 & 1.27 & 1.24 & 1.24 & 1.23 & 1.22 & 1.22 & 1.23 \\
$\mu(\neq snsssw)$ & 6.32 & 5.44 & 5.65 & 6.14 & 6.65 & 5.77 & 5.81 & 6.50 & 6.48 & 5.31 & 5.53 & 6.43 & 7.37 & 5.83 & 6.14 & 7.30 \\
$\sigma(\neq snsssw)$ & 3.07 & 2.83 & 2.97 & 3.04 & 3.07 & 2.90 & 2.92 & 3.08 & 2.93 & 2.60 & 2.70 & 2.98 & 3.37 & 3.02 & 3.26 & 3.39 \\ \hline
  \end{tabular}
  \caption{Sizes of tokens and words. Practically all sizes are greater for ELE. See subsection~\ref{subsec:tw2} for discussion and context.}
  \label{tab:sizesTokens}
\end{table*}

\begin{table*}
  \centering
    \footnotesize
\setlength{\tabcolsep}{.16667em}
  \begin{tabular}{|l|| c|c|c|c||  c|c|c|c||   c|c|c|c||   c|c|c|c|}\hline
-\-  & \multicolumn{4}{c|}{CPP} & \multicolumn{4}{c|}{LAD} & \multicolumn{4}{c|}{LAU} & \multicolumn{4}{c|}{ELE} \\ \hline
 & g. & p. & i. & h. &     g. & p. & i. & h. &    g. & p. & i. & h. &    g. & p. & i. & h. \\\hline
$|sents|$ & 106086 & 10154 & 17618 & 78309 & 113033 & 15581 & 15838 & 81608 & 111703 & 15822 & 19968 & 75926 & 325399 & 23835 & 36775 & 264794 \\ \hline
$\mu\left(\frac{chars}{sent}\right)$ & 118.31 & 148.63 & 125.02 & 112.87 & 110.52 & 125.69 & 116.16 & 106.54 & 105.15 & 120.64 & 107.55 & 101.27 & 117.67 & 126.06 & 128.01 & 115.48 \\
$\sigma\left(\frac{chars}{sent}\right)$ & 250.34 & 312.02 & 259.34 & 239.11 & 148.98 & 243.78 & 148.28 & 122.42 & 208.63 & 386.51 & 259.32 & 120.50 & 127.57 & 120.89 & 122.34 & 128.79 \\ \hline
$\mu\left(\frac{tokens}{sent}\right)$ & 26.80 & 34.06 & 28.91 & 25.38 & 23.79 & 27.04 & 25.03 & 22.93 & 23.20 & 26.40 & 23.98 & 22.33 & 24.68 & 26.78 & 27.29 & 24.13 \\
$\sigma\left(\frac{tokens}{sent}\right)$ & 64.74 & 81.47 & 64.30 & 62.36 & 33.44 & 51.90 & 29.21 & 29.40 & 38.11 & 51.39 & 54.91 & 27.88 & 34.48 & 27.38 & 29.18 & 35.69 \\ \hline
$\mu\left(\frac{kw}{sent}\right)$ & 13.88 & 16.09 & 12.99 & 13.80 & 15.15 & 15.76 & 15.67 & 14.94 & 14.11 & 14.39 & 13.98 & 14.08 & 17.03 & 17.76 & 17.88 & 16.84 \\
$\sigma\left(\frac{kw}{sent}\right)$ & 17.22 & 22.67 & 18.33 & 16.09 & 13.81 & 17.71 & 14.63 & 12.76 & 13.48 & 15.03 & 15.38 & 12.58 & 13.23 & 13.91 & 14.14 & 13.03 \\ \hline
$\mu\left(\frac{kwssnsw}{sent}\right)$ & 6.90 & 8.13 & 6.73 & 6.78 & 7.26 & 7.79 & 7.57 & 7.09 & 6.67 & 7.06 & 6.69 & 6.58 & 8.19 & 8.60 & 8.74 & 8.07 \\
$\sigma\left(\frac{kwssnsw}{sent}\right)$ & 10.72 & 14.17 & 11.76 & 9.92 & 7.79 & 11.11 & 7.84 & 6.95 & 7.54 & 8.92 & 9.71 & 6.49 & 6.59 & 7.05 & 7.13 & 6.46 \\ \hline
  \end{tabular}
  \caption{Sizes of sentences in characters and in tokens. Hubs produce the smallest sentences and, at the same time, present the lowest incidence of known words and of known words with synsets. See subsection~\ref{subsec:ss} for discussion and context.}
  \label{tab:sizesSents}
\end{table*}

\begin{table*}
  \centering
    \footnotesize
\setlength{\tabcolsep}{.16667em}
  \begin{tabular}{|l|| c|c|c|c||  c|c|c|c||   c|c|c|c||   c|c|c|c|}\hline
-\-  & \multicolumn{4}{c|}{CPP} & \multicolumn{4}{c|}{LAD} & \multicolumn{4}{c|}{LAU} & \multicolumn{4}{c|}{ELE} \\ \hline
 & g. & p. & i. & h. &     g. & p. & i. & h. &    g. & p. & i. & h. &    g. & p. & i. & h. \\\hline
$\mu\left(\frac{|chars|}{msg}\right)$ & 632.81 & 883.15 & 841.05 & 570.09 & 628.49 & 763.32 & 655.59 & 599.39 & 591.12 & 697.59 & 623.79 & 561.61 & 1934.43 & 1638.41 & 1796.38 & 1993.42 \\
$\sigma\left(\frac{|chars|}{msg}\right)$ & 1761.57 & 1247.79 & 3896.49 & 1101.55 & 836.23 & 1136.90 & 826.08 & 770.30 & 831.47 & 1194.85 & 982.59 & 686.75 & 2642.25 & 1737.49 & 1992.88 & 2819.96 \\\hline
$\mu\left(\frac{|tokens|}{msg}\right)$ & 143.35 & 202.36 & 194.09 & 128.28 & 135.99 & 164.49 & 141.88 & 129.81 & 131.37 & 153.18 & 139.27 & 125.01 & 406.39 & 347.64 & 383.28 & 417.36 \\
$\sigma\left(\frac{|tokens|}{msg}\right)$ & 444.20 & 287.17 & 940.83 & 304.37 & 178.11 & 237.80 & 172.03 & 165.98 & 173.89 & 213.52 & 212.91 & 152.35 & 557.29 & 365.05 & 435.87 & 593.08 \\\hline
$\mu\left(\frac{|sents|}{msg}\right)$ & 5.71 & 6.39 & 7.09 & 5.40 & 6.12 & 6.55 & 6.11 & 6.04 & 6.08 & 6.23 & 6.23 & 6.01 & 17.22 & 13.74 & 14.79 & 18.05 \\
$\sigma\left(\frac{|sents|}{msg}\right)$ & 16.36 & 6.29 & 41.76 & 6.55 & 6.75 & 7.51 & 6.67 & 6.61 & 6.58 & 8.03 & 6.87 & 6.18 & 23.97 & 14.06 & 17.01 & 25.80 \\\hline
  \end{tabular}
  \caption{Mean and standard deviation of message sizes. Greater size of messages from ELE list reflects domain of interest, as does its hubsi sector, which produces the largest texts. See subsection~\ref{subsec:mm} for discussion and context.}
  \label{tab:sizesMsgs}
\end{table*}

\renewcommand{\arraystretch}{1.3}
\begin{table*}
  \centering
    \small
\setlength{\tabcolsep}{.26667em}
  \begin{tabular}{|l|| c|c|c|c||  c|c|c|c||   c|c|c|c||   c|c|c|c|}\hline
-\-  & \multicolumn{4}{c|}{CPP} & \multicolumn{4}{c|}{LAD} & \multicolumn{4}{c|}{LAU} & \multicolumn{4}{c|}{ELE} \\ \hline
 & g. & p. & i. & h. &     g. & p. & i. & h. &    g. & p. & i. & h. &    g. & p. & i. & h. \\\hline
NN & 28.17 & 30.38 & 31.13 & 27.19 & 26.68 & 29.29 & 26.98 & 26.08 & 26.64 & 29.87 & 28.03 & 25.58 & 24.68 & 25.54 & 25.35 & 24.50\\
NNS & 2.51 & 2.32 & 2.56 & 2.53 & 2.82 & 2.97 & 2.92 & 2.76 & 2.63 & 2.63 & 2.65 & 2.63 & 4.41 & 4.56 & 4.61 & 4.36\\
NNP & 0.72 & 0.75 & 1.03 & 0.65 & 0.70 & 1.10 & 0.74 & 0.61 & 0.90 & 0.94 & 0.94 & 0.88 & 0.76 & 1.13 & 1.04 & 0.69\\
NNPS & 0.01 & 0.01 & 0.00 & 0.01 & 0.01 & 0.03 & 0.02 & 0.01 & 0.01 & 0.01 & 0.02 & 0.01 & 0.03 & 0.05 & 0.02 & 0.03\\\hline
+ & 31.41 & 33.46 & 34.73 & 30.38 & 30.21 & 33.39 & 30.65 & 29.47 & 30.18 & 33.45 & 31.63 & 29.10 & 29.88 & 31.29 & 31.02 & 29.58\\\hline\hline
JJ & 4.83 & 4.60 & 4.72 & 4.89 & 5.05 & 5.03 & 5.00 & 5.06 & 4.65 & 4.46 & 4.42 & 4.75 & 5.19 & 5.11 & 5.24 & 5.19\\
JJR & 0.45 & 0.37 & 0.38 & 0.48 & 0.47 & 0.43 & 0.48 & 0.48 & 0.45 & 0.36 & 0.40 & 0.48 & 0.66 & 0.71 & 0.73 & 0.65\\
JJS & 0.17 & 0.15 & 0.14 & 0.17 & 0.25 & 0.22 & 0.26 & 0.26 & 0.25 & 0.22 & 0.22 & 0.26 & 0.38 & 0.41 & 0.46 & 0.37\\
RB & 6.43 & 5.29 & 5.73 & 6.76 & 6.55 & 5.41 & 6.30 & 6.83 & 6.60 & 5.74 & 6.11 & 6.91 & 5.78 & 5.27 & 5.34 & 5.89\\
RBR & 0.11 & 0.08 & 0.09 & 0.12 & 0.12 & 0.10 & 0.12 & 0.12 & 0.11 & 0.07 & 0.09 & 0.12 & 0.16 & 0.14 & 0.16 & 0.16\\
RBS & 0.02 & 0.01 & 0.01 & 0.02 & 0.03 & 0.02 & 0.03 & 0.03 & 0.02 & 0.01 & 0.02 & 0.02 & 0.04 & 0.05 & 0.04 & 0.04\\
RP & 0.35 & 0.30 & 0.27 & 0.37 & 0.39 & 0.36 & 0.43 & 0.39 & 0.50 & 0.43 & 0.50 & 0.52 & 0.26 & 0.30 & 0.25 & 0.26\\\hline
+ & 12.36 & 10.79 & 11.34 & 12.82 & 12.86 & 11.59 & 12.61 & 13.17 & 12.58 & 11.29 & 11.76 & 13.08 & 12.47 & 12.00 & 12.23 & 12.55\\\hline\hline
VB & 6.25 & 6.24 & 6.31 & 6.25 & 5.90 & 5.72 & 5.91 & 5.94 & 5.89 & 5.98 & 5.92 & 5.86 & 5.22 & 5.27 & 5.06 & 5.24\\
VBZ & 3.94 & 3.89 & 3.80 & 3.97 & 3.97 & 3.60 & 3.87 & 4.07 & 3.77 & 3.48 & 3.58 & 3.88 & 4.16 & 3.79 & 4.14 & 4.20\\
VBP & 3.17 & 3.07 & 3.17 & 3.18 & 2.84 & 2.63 & 2.86 & 2.87 & 3.23 & 2.93 & 3.11 & 3.32 & 2.68 & 2.64 & 2.67 & 2.68\\
VBN & 2.00 & 2.14 & 2.06 & 1.97 & 1.78 & 1.85 & 1.93 & 1.74 & 1.74 & 1.78 & 1.75 & 1.72 & 1.87 & 2.02 & 1.80 & 1.86\\
VBD & 1.52 & 1.64 & 1.49 & 1.50 & 1.38 & 1.43 & 1.47 & 1.35 & 1.71 & 1.64 & 1.69 & 1.74 & 1.49 & 1.41 & 1.48 & 1.51\\
VBG & 1.50 & 1.66 & 1.41 & 1.50 & 1.57 & 1.69 & 1.58 & 1.54 & 1.66 & 1.76 & 1.71 & 1.63 & 1.51 & 1.59 & 1.55 & 1.50\\
MD & 2.20 & 1.78 & 2.09 & 2.28 & 2.31 & 2.07 & 2.20 & 2.38 & 2.16 & 1.99 & 2.07 & 2.22 & 2.44 & 2.25 & 2.16 & 2.51\\\hline
+ & 20.58 & 20.42 & 20.32 & 20.66 & 19.75 & 18.99 & 19.82 & 19.89 & 20.16 & 19.55 & 19.84 & 20.37 & 19.37 & 18.98 & 18.87 & 19.48\\\hline\hline
IN & 12.60 & 12.49 & 12.08 & 12.73 & 12.15 & 12.17 & 12.18 & 12.14 & 11.97 & 11.70 & 11.99 & 12.02 & 13.11 & 13.18 & 13.06 & 13.12\\
DT & 10.76 & 10.96 & 10.33 & 10.82 & 10.81 & 10.56 & 10.81 & 10.86 & 10.45 & 10.28 & 10.48 & 10.48 & 11.57 & 11.77 & 11.55 & 11.55\\
PRP & 3.62 & 2.83 & 3.02 & 3.87 & 4.06 & 3.40 & 3.85 & 4.25 & 4.34 & 3.48 & 3.95 & 4.63 & 3.56 & 3.06 & 3.21 & 3.66\\
PRP\$ & 0.73 & 0.85 & 0.56 & 0.75 & 0.99 & 1.01 & 1.00 & 0.99 & 1.15 & 1.15 & 1.16 & 1.14 & 0.97 & 0.96 & 1.04 & 0.96\\
PDT & 0.08 & 0.08 & 0.07 & 0.08 & 0.08 & 0.07 & 0.09 & 0.08 & 0.08 & 0.06 & 0.09 & 0.08 & 0.10 & 0.08 & 0.12 & 0.10\\
TO & 2.93 & 2.94 & 2.87 & 2.94 & 3.16 & 3.19 & 3.20 & 3.14 & 3.13 & 3.15 & 3.20 & 3.10 & 2.92 & 2.95 & 2.91 & 2.92\\
CC & 2.77 & 2.97 & 2.54 & 2.79 & 3.52 & 3.55 & 3.56 & 3.50 & 3.61 & 3.63 & 3.66 & 3.59 & 3.03 & 2.94 & 3.16 & 3.03\\
WRB & 0.58 & 0.68 & 0.56 & 0.56 & 0.59 & 0.51 & 0.55 & 0.61 & 0.59 & 0.60 & 0.58 & 0.58 & 0.64 & 0.57 & 0.58 & 0.66\\
WDT & 0.54 & 0.53 & 0.55 & 0.54 & 0.54 & 0.48 & 0.49 & 0.56 & 0.48 & 0.42 & 0.45 & 0.50 & 0.60 & 0.56 & 0.59 & 0.61\\
WP & 0.32 & 0.28 & 0.29 & 0.33 & 0.44 & 0.35 & 0.41 & 0.46 & 0.47 & 0.42 & 0.41 & 0.49 & 0.58 & 0.50 & 0.50 & 0.60\\
WP\$ & 0.00 & 0.00 & 0.00 & 0.00 & 0.00 & 0.00 & 0.00 & 0.00 & 0.00 & 0.00 & 0.00 & 0.00 & 0.02 & 0.01 & 0.02 & 0.02\\\hline
+ & 34.93 & 34.62 & 32.87 & 35.42 & 36.34 & 35.29 & 36.13 & 36.61 & 36.26 & 34.89 & 35.98 & 36.62 & 37.12 & 36.60 & 36.74 & 37.22\\\hline\hline
CD & 0.38 & 0.37 & 0.36 & 0.38 & 0.44 & 0.41 & 0.44 & 0.45 & 0.42 & 0.37 & 0.43 & 0.43 & 0.79 & 0.78 & 0.81 & 0.79\\
EX & 0.27 & 0.28 & 0.29 & 0.27 & 0.33 & 0.29 & 0.28 & 0.35 & 0.33 & 0.33 & 0.29 & 0.34 & 0.35 & 0.34 & 0.31 & 0.36\\
UH & 0.07 & 0.04 & 0.08 & 0.07 & 0.04 & 0.03 & 0.04 & 0.04 & 0.04 & 0.03 & 0.04 & 0.05 & 0.01 & 0.01 & 0.01 & 0.01\\
FW & 0.01 & 0.03 & 0.00 & 0.00 & 0.02 & 0.02 & 0.03 & 0.02 & 0.03 & 0.09 & 0.02 & 0.01 & 0.00 & 0.01 & 0.00 & 0.00\\\hline
+ & 0.72 & 0.71 & 0.74 & 0.72 & 0.84 & 0.74 & 0.78 & 0.86 & 0.82 & 0.82 & 0.79 & 0.83 & 1.16 & 1.14 & 1.13 & 1.17\\\hline\hline
  \end{tabular}
  \caption{Incidence of Brown Tags. Used Brill tagger with $\approx 85\%$ of correctly identified tags on the Brown Corpus. Most explicit is the peripheral incidence of nouns and hubs incidence of adjectives, adverbs and verbs. See subsection~\ref{subsec:pos} for discussion and context.}
  \label{tab:pos}
\end{table*}


\begin{table}
  \centering
    \small
\setlength{\tabcolsep}{.26667em}
  \begin{tabular}{|l|| c|c|c|}\hline
list$\setminus$measure & H-P & H-I & I-P \\\hline
CPP & 5.58 & 2.54 & 7.82 \\\hline
LAD & 7.67 & 2.07 & 8.35 \\\hline
LAU & 6.23 & 1.63 & 5.98 \\\hline
ELE & 3.42 & 0.77 & 2.81 \\\hline
  \end{tabular}
  \caption{Kolmogorov $c(\alpha)$ values for substantives. See subsection~\ref{subsec:di} for discussion and directions.}
  \label{tab:kolSub}
\end{table}

\begin{table}
  \centering
    \small
\setlength{\tabcolsep}{.06667em}
  \begin{tabular}{|l|| c|c|c|c|c|c|}\hline
& CPP-LAD & CPP-LAU & CPP-ELE & LAD-LAU & LAD-ELE & LAU-ELE \\\hline
P & 1.35 & 4.05 & 5.80 & 3.00 & 5.41 & 4.94 \\\hline
I & 1.27 & 0.78 & 4.01 & 0.84 & 3.84 & 3.94 \\\hline
H & 0.98 & 1.94 & 3.17 & 1.32 & 3.82 & 4.47 \\\hline
  \end{tabular}
  \caption{Kolmogorov $c(\alpha)$ values for substantives. Comparrison of the same sector between lists, each author is an observation. See subsection~\ref{subsec:di} for discussion and directions.}
  \label{tab:kolSubInter}
\end{table}

\begin{table}
  \centering
    \small
\setlength{\tabcolsep}{.26667em}
  \begin{tabular}{|l|| c|c|c|}\hline
list$\setminus$measure & H-P & H-I & I-P \\\hline
CPP & 2.76 & 2.33 & 0.25 \\\hline
LAD & 4.22 & 2.88 & 1.02 \\\hline
LAU & 4.30 & 2.45 & 1.34 \\\hline
ELE & 4.77 & 1.69 & 2.86 \\\hline
  \end{tabular}
  \caption{Kolmogorov $c(\alpha)$ values for adjectives. See subsection~\ref{subsec:di} for discussion and directions.}
  \label{tab:kolAdj}
\end{table}

\begin{table}
  \centering
    \small
\setlength{\tabcolsep}{.06667em}
  \begin{tabular}{|l|| c|c|c|c|c|c|}\hline
 & CPP-LAD & CPP-LAU & CPP-ELE & LAD-LAU & LAD-ELE & LAU-ELE \\\hline
P & 0.44 & 0.34 & 2.57 & 0.20 & 2.32 & 2.37 \\\hline
I & 0.74 & 0.99 & 3.72 & 0.32 & 3.37 & 3.10 \\\hline
H & 0.26 & 0.32 & 3.72 & 0.29 & 4.36 & 4.24 \\\hline
  \end{tabular}
  \caption{Kolmogorov $c(\alpha)$ values for adjectives. Comparrison of the same sector between lists, each author is an observation. See subsection~\ref{subsec:di} for discussion and directions.}
  \label{tab:kolAdjInter}
\end{table}

\begin{table}
  \centering
    \small
\setlength{\tabcolsep}{.26667em}
  \begin{tabular}{|l|| c|c|c|}\hline
list$\setminus$measure & H-P & H-I & I-P \\\hline
CPP & 7.01 & 4.89 & 7.95 \\\hline
LAD & 9.82 & 6.13 & 8.58 \\\hline
LAU & 9.66 & 5.44 & 7.45 \\\hline
ELE & 5.78 & 2.84 & 4.69 \\\hline
  \end{tabular}
  \caption{Kolmogorov $c(\alpha)$ values for stopwords. See subsection~\ref{subsec:di} for discussion and directions.}
  \label{tab:kolSw}
\end{table}

\begin{table}
  \centering
    \small
\setlength{\tabcolsep}{.06667em}
  \begin{tabular}{|l|| c|c|c|c|c|c|}\hline
 & CPP-LAD & CPP-LAU & CPP-ELE & LAD-LAU & LAD-ELE & LAU-ELE \\\hline
P & 3.31 & 3.26 & 6.68 & 0.57 & 5.36 & 5.41 \\\hline
I & 1.45 & 1.08 & 5.16 & 0.91 & 5.00 & 4.92 \\\hline
H & 0.98 & 0.68 & 4.35 & 1.05 & 4.73 & 5.01 \\\hline
  \end{tabular}
  \caption{Kolmogorov $c(\alpha)$ values for stopwords. Comparrison of the same sector between lists, each author is an observation. See subsection~\ref{subsec:di} for discussion and directions.}
  \label{tab:kolSwInter}
\end{table}

\begin{table}
  \centering
    \small
\setlength{\tabcolsep}{.26667em}
  \begin{tabular}{|l|| c|c|c|}\hline
list$\setminus$measure & H-P & H-I & I-P \\\hline
CPP & 1.53 & 0.89 & 1.45 \\\hline
LAD & 2.32 & 0.97 & 2.09 \\\hline
LAU & 2.10 & 0.78 & 1.68 \\\hline
ELE & 1.51 & 1.32 & 1.15 \\\hline
  \end{tabular}
  \caption{Kolmogorov $c(\alpha)$ values for punctuations/char. See subsection~\ref{subsec:di} for discussion and directions.}
  \label{tab:kolPct}
\end{table}

\begin{table}
  \centering
    \small
\setlength{\tabcolsep}{.06667em}
  \begin{tabular}{|l|| c|c|c|c|c|c|}\hline
 & CPP-LAD & CPP-LAU & CPP-ELE & LAD-LAU & LAD-ELE & LAU-ELE \\\hline
P & 5.74 & 4.88 & 8.28 & 2.23 & 5.37 & 6.60 \\\hline
I & 3.23 & 2.49 & 4.16 & 0.96 & 3.40 & 3.51 \\\hline
H & 2.49 & 1.87 & 4.02 & 1.36 & 3.05 & 3.71 \\\hline
  \end{tabular}
  \caption{Kolmogorov $c(\alpha)$ values for punctuations/char. Comparrison of the same sector between lists, each author is an observation. See subsection~\ref{subsec:di} for discussion and directions.}
  \label{tab:kolPctInter}
\end{table}

\begin{table*}
  \centering
    \footnotesize
\setlength{\tabcolsep}{.26667em}
  \begin{tabular}{|l|| c|c|c|c||  c|c|c|c||   c|c|c|c||   c|c|c|c|}\hline
-\-  & \multicolumn{4}{c|}{CPP} & \multicolumn{4}{c|}{LAD} & \multicolumn{4}{c|}{LAU} & \multicolumn{4}{c|}{ELE} \\ \hline
 & g. & p. & i. & h. &     g. & p. & i. & h. &    g. & p. & i. & h. &    g. & p. & i. & h. \\\hline
$d$ - $d_i$ & 0.9972 & 0.8188 & 0.9477 & 1.0162 & 0.9927 & 0.8268 & 0.9028 & 0.9952 & 0.9906 & 0.8045 & 0.7900 & 0.9948 & 0.9752 & 0.8331 & 0.9057 & 0.9791 \\ \hline
$d$ - $d_o$ & 0.9932 & 0.8517 & 0.9381 & 1.0126 & 0.9907 & 0.8697 & 0.8696 & 0.9904 & 0.9842 & 0.8624 & 0.6289 & 0.9798 & 0.9529 & 0.8760 & 0.5669 & 0.8636 \\ \hline
$d$ - $s$ & 0.9572 & 0.9167 & 0.8598 & 0.9835 & 0.9525 & 0.9685 & 0.8991 & 0.9592 & 0.9741 & 0.9715 & 0.9066 & 0.9811 & 0.9010 & 0.9557 & 0.5912 & 0.8480 \\ \hline
$d$ - $s_i$ & 0.9539 & 0.7715 & 0.8329 & 0.9893 & 0.9420 & 0.8068 & 0.8366 & 0.9451 & 0.9628 & 0.7926 & 0.7378 & 0.9748 & 0.8695 & 0.8142 & 0.6811 & 0.8588 \\ \hline
$d$ - $s_o$ & 0.9547 & 0.7662 & 0.7574 & 0.9692 & 0.9516 & 0.8406 & 0.7230 & 0.9572 & 0.9668 & 0.8452 & 0.4615 & 0.9575 & 0.8785 & 0.8218 & 0.0913 & 0.7152 \\ \hline
$d$ - $bc$ & 0.9698 & 0.5773 & 0.7471 & 0.9921 & 0.9488 & 0.4707 & 0.6327 & 0.9603 & 0.9561 & 0.4917 & 0.5860 & 0.9635 & 0.9277 & 0.7236 & 0.8108 & 0.9057 \\ \hline
$d$ - $triangles$ & 0.9716 & 0.7773 & 0.9342 & 0.9742 & 0.9789 & 0.8035 & 0.8644 & 0.9981 & 0.9752 & 0.7987 & 0.8110 & 0.9856 & 0.9889 & 0.9213 & 0.9455 & 0.9969 \\ \hline
$d_i$ - $d_o$ & 0.9787 & 0.3936 & 0.7615 & 1.0031 & 0.9647 & 0.4389 & 0.5605 & 0.9473 & 0.9481 & 0.3905 & 0.0068 & 0.9283 & 0.8524 & 0.4521 & 0.0925 & 0.6616 \\ \hline
$d_i$ - $s$ & 0.9595 & 0.7348 & 0.8066 & 0.9757 & 0.9529 & 0.7929 & 0.7852 & 0.9548 & 0.9700 & 0.7766 & 0.6592 & 0.9646 & 0.8809 & 0.7564 & 0.4000 & 0.7913 \\ \hline
$d_i$ - $s_i$ & 0.9601 & 0.9315 & 0.8802 & 0.9838 & 0.9529 & 0.9675 & 0.9298 & 0.9565 & 0.9749 & 0.9744 & 0.9613 & 0.9800 & 0.8918 & 0.9633 & 0.8697 & 0.8654 \\ \hline
$d_i$ - $s_o$ & 0.9523 & 0.3664 & 0.6185 & 0.9587 & 0.9408 & 0.4299 & 0.4345 & 0.9357 & 0.9433 & 0.4010 & -0.1314 & 0.9136 & 0.8045 & 0.4266 & -0.2781 & 0.5651 \\ \hline
$d_i$ - $bc$ & 0.9780 & 0.4765 & 0.7036 & 0.9970 & 0.9453 & 0.4052 & 0.6128 & 0.9463 & 0.9612 & 0.4369 & 0.5146 & 0.9617 & 0.9283 & 0.7161 & 0.7301 & 0.8838 \\ \hline
$d_i$ - $triangles$ & 0.9599 & 0.5910 & 0.8621 & 0.9634 & 0.9713 & 0.6370 & 0.7540 & 0.9780 & 0.9683 & 0.5342 & 0.5127 & 0.9636 & 0.9526 & 0.6953 & 0.7154 & 0.9152 \\ \hline
$d_o$ - $s$ & 0.9413 & 0.7934 & 0.8052 & 0.9866 & 0.9338 & 0.8477 & 0.8041 & 0.9329 & 0.9505 & 0.8405 & 0.6325 & 0.9587 & 0.8488 & 0.8652 & 0.5383 & 0.7486 \\ \hline
$d_o$ - $s_o$ & 0.9457 & 0.8894 & 0.8084 & 0.9770 & 0.9455 & 0.9626 & 0.8712 & 0.9509 & 0.9682 & 0.9656 & 0.9106 & 0.9759 & 0.8999 & 0.9401 & 0.6816 & 0.8175 \\ \hline
$d_o$ - $bc$ & 0.9452 & 0.4872 & 0.6967 & 0.9749 & 0.9346 & 0.3945 & 0.4971 & 0.9451 & 0.9211 & 0.3879 & 0.2974 & 0.9207 & 0.8457 & 0.5296 & 0.4184 & 0.7291 \\ \hline
$d_o$ - $triangles$ & 0.9756 & 0.7021 & 0.8903 & 0.9825 & 0.9686 & 0.7215 & 0.7741 & 0.9888 & 0.9550 & 0.7799 & 0.6628 & 0.9710 & 0.9506 & 0.8638 & 0.7274 & 0.9073 \\ \hline
$s$ - $s_i$ & 0.9985 & 0.7926 & 0.9230 & 1.0162 & 0.9951 & 0.8225 & 0.8694 & 1.0002 & 0.9928 & 0.8061 & 0.7107 & 0.9970 & 0.9799 & 0.7942 & 0.5159 & 0.9919 \\ \hline
$s$ - $s_o$ & 0.9971 & 0.8764 & 0.9345 & 1.0146 & 0.9942 & 0.8785 & 0.8727 & 0.9987 & 0.9891 & 0.8795 & 0.6317 & 0.9898 & 0.9631 & 0.9069 & 0.7149 & 0.9383 \\ \hline
$s$ - $triangles$ & 0.9298 & 0.6961 & 0.8118 & 0.9518 & 0.9616 & 0.7829 & 0.7600 & 0.9471 & 0.9741 & 0.7713 & 0.7583 & 0.9613 & 0.8933 & 0.8715 & 0.5878 & 0.7889 \\ \hline
$s_i$ - $s_o$ & 0.9886 & 0.3980 & 0.7088 & 1.0062 & 0.9764 & 0.4482 & 0.5053 & 0.9732 & 0.9617 & 0.4246 & -0.1088 & 0.9506 & 0.8801 & 0.4528 & -0.2813 & 0.8041 \\ \hline
$s_i$ - $triangles$ & 0.9227 & 0.5365 & 0.7822 & 0.9552 & 0.9492 & 0.6210 & 0.6608 & 0.9281 & 0.9625 & 0.5225 & 0.4700 & 0.9455 & 0.8553 & 0.6719 & 0.4366 & 0.7793 \\ \hline
$s_o$ - $triangles$ & 0.9321 & 0.6209 & 0.7191 & 0.9410 & 0.9626 & 0.7058 & 0.6569 & 0.9504 & 0.9672 & 0.7570 & 0.5478 & 0.9505 & 0.8799 & 0.7989 & 0.2907 & 0.6970 \\ \hline
$bc$ - $triangles$ & 0.9055 & 0.4769 & 0.6933 & 0.9031 & 0.9555 & 0.2694 & 0.4095 & 0.9467 & 0.9409 & 0.2329 & 0.2844 & 0.9129 & 0.9255 & 0.7423 & 0.7459 & 0.8793 \\ \hline
$IC$ - $IP$ & -1.0010 & -1.0012 & 0.0000 & 0.0000 & -1.0008 & -1.0011 & 0.0000 & 0.0000 & -1.0008 & -1.0011 & 0.0000 & 0.0000 & -1.0033 & -1.0045 & 0.0000 & 0.0000 \\ \hline
  \end{tabular}
  \caption{Correlation of topological measures. See subsection~\ref{subsec:cor} for discussion and directions.}
  \label{tab:corTop}
\end{table*}

\clearpage
\begin{longtable*}{|p{2cm}||c|c|c|c|| c|c|c|c|| c|c|c|c|| c|c|c|c|}\hline
  \centering
    \small
-\-  & \multicolumn{4}{c|}{CPP} & \multicolumn{4}{c|}{LAD} & \multicolumn{4}{c|}{LAU} & \multicolumn{4}{c|}{ELE} \\ \hline\hline
 & g. & p. & i. & h. &     g. & p. & i. & h. &    g. & p. & i. & h. &    g. & p. & i. & h. \\\hline \endfirsthead
-\-  & \multicolumn{4}{c|}{CPP} & \multicolumn{4}{c|}{LAD} & \multicolumn{4}{c|}{LAU} & \multicolumn{4}{c|}{ELE} \\ \hline\hline
 & g. & p. & i. & h. &     g. & p. & i. & h. &    g. & p. & i. & h. &    g. & p. & i. & h. \\\hline \endhead
nc-nt & 1.000 & 0.978 & 0.992 & 1.018 & 1.000 & 0.994 & 1.001 & 1.008 & 1.000 & 0.940 & 0.995 & 1.007 & 1.003 & 1.002 & 1.025 & 1.025 \\ \hline
np/(nc-ne)-ntp/nt & 0.934 & 0.936 & 0.930 & 0.983 & 0.893 & 0.890 & 0.956 & 0.932 & 0.933 & 0.934 & 0.961 & 0.911 & 0.963 & 0.963 & 1.012 & 0.940 \\ \hline
nt-ntd & 0.927 & 0.870 & 0.837 & 0.988 & 0.943 & 0.918 & 0.954 & 0.967 & 0.956 & 0.921 & 0.947 & 0.967 & 0.807 & 0.946 & 0.948 & 0.923 \\ \hline
Nwss/Nkw-Nwss\_/Nkw\_ & 0.805 & 0.862 & -0.107 & -0.401 & 0.877 & 0.922 & 0.882 & -0.040 & 0.880 & 0.920 & 0.958 & -0.010 & 0.824 & 0.869 & -0.319 & -0.394 \\ \hline
Nwsw/Nkw-Nwsssw/Nwss & 0.890 & 0.882 & 0.960 & 0.995 & 0.903 & 0.899 & 0.935 & 0.941 & 0.915 & 0.912 & 0.956 & 0.912 & 0.931 & 0.931 & 0.989 & 0.884 \\ \hline
mtkw-mtkwnsw & 0.855 & 0.868 & 0.386 & 0.388 & 0.941 & 0.943 & 0.943 & 0.769 & 0.944 & 0.943 & 0.971 & 0.826 & 0.937 & 0.940 & 0.927 & 0.758 \\ \hline
mtkw-mtkwnsw\_ & 0.849 & 0.878 & 0.447 & 0.125 & 0.915 & 0.939 & 0.929 & 0.426 & 0.913 & 0.935 & 0.951 & 0.409 & 0.823 & 0.904 & 0.622 & 0.238 \\ \hline
mtkw-mtams & 0.855 & 0.867 & 0.434 & 0.450 & 0.942 & 0.944 & 0.946 & 0.786 & 0.945 & 0.944 & 0.974 & 0.837 & 0.951 & 0.954 & 0.940 & 0.785 \\ \hline
mtkw-mtams\_ & 0.846 & 0.873 & 0.483 & 0.120 & 0.916 & 0.939 & 0.930 & 0.428 & 0.913 & 0.935 & 0.953 & 0.404 & 0.842 & 0.922 & 0.620 & 0.244 \\ \hline
dtkw-dtkw\_ & 0.962 & 0.969 & 0.739 & 0.612 & 0.979 & 0.984 & 0.942 & 0.660 & 0.977 & 0.982 & 0.966 & 0.605 & 0.963 & 0.972 & 0.786 & 0.399 \\ \hline
dtkw-mtkwnsw & 0.851 & 0.854 & 0.788 & 0.814 & 0.927 & 0.926 & 0.942 & 0.920 & 0.919 & 0.916 & 0.966 & 0.836 & 0.938 & 0.941 & 0.956 & 0.924 \\ \hline
dtkw-dtkwnsw & 0.903 & 0.904 & 0.890 & 0.833 & 0.936 & 0.936 & 0.952 & 0.902 & 0.941 & 0.940 & 0.975 & 0.873 & 0.938 & 0.944 & 0.810 & 0.902 \\ \hline
dtkw-mtkwnsw\_ & 0.833 & 0.845 & 0.778 & 0.564 & 0.908 & 0.923 & 0.920 & 0.457 & 0.903 & 0.916 & 0.935 & 0.478 & 0.837 & 0.914 & 0.674 & 0.399 \\ \hline
dtkw-dtkwnsw\_ & 0.879 & 0.888 & 0.620 & 0.507 & 0.917 & 0.923 & 0.917 & 0.598 & 0.923 & 0.927 & 0.953 & 0.564 & 0.924 & 0.942 & 0.655 & 0.358 \\ \hline
dtkw-mtams & 0.848 & 0.850 & 0.825 & 0.815 & 0.929 & 0.929 & 0.945 & 0.921 & 0.921 & 0.918 & 0.970 & 0.848 & 0.937 & 0.939 & 0.965 & 0.942 \\ \hline
dtkw-dtams & 0.887 & 0.887 & 0.882 & 0.805 & 0.928 & 0.928 & 0.948 & 0.902 & 0.936 & 0.935 & 0.972 & 0.872 & 0.930 & 0.935 & 0.778 & 0.892 \\ \hline
dtkw-mtams\_ & 0.826 & 0.838 & 0.784 & 0.555 & 0.910 & 0.925 & 0.921 & 0.457 & 0.904 & 0.917 & 0.937 & 0.475 & 0.846 & 0.921 & 0.673 & 0.410 \\ \hline
dtkw-dtams\_ & 0.867 & 0.875 & 0.610 & 0.506 & 0.911 & 0.916 & 0.914 & 0.607 & 0.920 & 0.923 & 0.952 & 0.577 & 0.921 & 0.937 & 0.661 & 0.385 \\ \hline
mtkw\_-mtkwnsw\_ & 0.871 & 0.907 & 0.912 & 1.007 & 0.913 & 0.941 & 0.964 & 0.993 & 0.916 & 0.941 & 0.976 & 0.993 & 0.943 & 0.946 & 1.002 & 1.021 \\ \hline
mtkw\_-mtams\_ & 0.863 & 0.899 & 0.901 & 1.008 & 0.912 & 0.941 & 0.964 & 0.993 & 0.915 & 0.940 & 0.976 & 0.995 & 0.932 & 0.934 & 1.002 & 1.019 \\ \hline
mtkw\_-mtsw\_ & 0.823 & 0.773 & 0.753 & 0.743 & 0.889 & 0.861 & 0.876 & 0.790 & 0.904 & 0.879 & 0.945 & 0.839 & 0.933 & 0.935 & 0.650 & 0.748 \\ \hline
mtkw\_-mtsw2\_ & 0.838 & 0.768 & 0.774 & 0.897 & 0.901 & 0.867 & 0.871 & 0.856 & 0.906 & 0.871 & 0.941 & 0.860 & 0.944 & 0.946 & 0.744 & 0.844 \\ \hline
dtkw\_-mtkwnsw & 0.821 & 0.829 & 0.598 & 0.598 & 0.915 & 0.917 & 0.908 & 0.632 & 0.905 & 0.903 & 0.964 & 0.563 & 0.908 & 0.914 & 0.686 & 0.321 \\ \hline
dtkw\_-dtkwnsw & 0.896 & 0.901 & 0.687 & 0.518 & 0.940 & 0.941 & 0.942 & 0.625 & 0.939 & 0.942 & 0.950 & 0.540 & 0.928 & 0.936 & 0.736 & 0.537 \\ \hline
dtkw\_-mtkwnsw\_ & 0.851 & 0.860 & 0.765 & 0.752 & 0.920 & 0.929 & 0.935 & 0.696 & 0.920 & 0.922 & 0.974 & 0.823 & 0.849 & 0.912 & 0.639 & 0.478 \\ \hline
dtkw\_-dtkwnsw\_ & 0.929 & 0.930 & 0.935 & 0.992 & 0.951 & 0.951 & 0.993 & 0.989 & 0.959 & 0.957 & 1.002 & 0.993 & 0.971 & 0.978 & 0.971 & 1.004 \\ \hline
dtkw\_-mtams & 0.822 & 0.829 & 0.641 & 0.623 & 0.917 & 0.919 & 0.909 & 0.625 & 0.907 & 0.905 & 0.965 & 0.570 & 0.906 & 0.911 & 0.693 & 0.345 \\ \hline
dtkw\_-dtams & 0.877 & 0.882 & 0.679 & 0.531 & 0.933 & 0.934 & 0.939 & 0.628 & 0.934 & 0.937 & 0.946 & 0.534 & 0.922 & 0.929 & 0.708 & 0.523 \\ \hline
dtkw\_-mtams\_ & 0.845 & 0.853 & 0.770 & 0.752 & 0.922 & 0.931 & 0.936 & 0.693 & 0.921 & 0.924 & 0.974 & 0.824 & 0.857 & 0.918 & 0.656 & 0.506 \\ \hline
dtkw\_-dtams\_ & 0.914 & 0.914 & 0.929 & 0.994 & 0.945 & 0.944 & 0.991 & 0.989 & 0.955 & 0.953 & 1.000 & 0.993 & 0.968 & 0.973 & 0.978 & 1.007 \\ \hline
mtkwnsw-mtkwnsw\_ & 0.940 & 0.968 & 0.754 & 0.633 & 0.972 & 0.990 & 0.959 & 0.567 & 0.968 & 0.986 & 0.980 & 0.609 & 0.904 & 0.968 & 0.767 & 0.534 \\ \hline
mtkwnsw-mtams & 0.985 & 0.985 & 0.986 & 0.994 & 0.998 & 0.998 & 1.001 & 1.004 & 0.997 & 0.997 & 1.005 & 1.003 & 0.999 & 1.000 & 1.025 & 1.014 \\ \hline
mtkwnsw-mtams\_ & 0.930 & 0.957 & 0.734 & 0.620 & 0.970 & 0.988 & 0.958 & 0.565 & 0.965 & 0.982 & 0.980 & 0.603 & 0.911 & 0.971 & 0.761 & 0.539 \\ \hline
dtkwnsw-dtkwnsw\_ & 0.959 & 0.968 & 0.661 & 0.459 & 0.979 & 0.985 & 0.939 & 0.579 & 0.973 & 0.981 & 0.941 & 0.520 & 0.943 & 0.967 & 0.678 & 0.497 \\ \hline
dtkwnsw-dtams & 0.988 & 0.988 & 0.993 & 1.007 & 0.994 & 0.994 & 1.001 & 1.005 & 0.992 & 0.992 & 1.003 & 1.002 & 0.997 & 0.999 & 1.019 & 1.012 \\ \hline
dtkwnsw-dtams\_ & 0.951 & 0.960 & 0.645 & 0.441 & 0.973 & 0.978 & 0.937 & 0.573 & 0.966 & 0.973 & 0.942 & 0.533 & 0.943 & 0.963 & 0.695 & 0.527 \\ \hline
mtkwnsw\_-mtams & 0.938 & 0.965 & 0.772 & 0.630 & 0.969 & 0.987 & 0.955 & 0.562 & 0.966 & 0.983 & 0.978 & 0.606 & 0.891 & 0.959 & 0.768 & 0.533 \\ \hline
mtkwnsw\_-mtams\_ & 0.993 & 0.992 & 1.000 & 1.018 & 0.998 & 0.999 & 1.004 & 1.007 & 0.998 & 0.998 & 1.006 & 1.007 & 1.001 & 1.001 & 1.027 & 1.024 \\ \hline
dtkwnsw\_-dtams & 0.946 & 0.954 & 0.648 & 0.481 & 0.973 & 0.978 & 0.935 & 0.579 & 0.965 & 0.973 & 0.938 & 0.514 & 0.940 & 0.963 & 0.666 & 0.490 \\ \hline
dtkwnsw\_-dtams\_ & 0.990 & 0.990 & 0.991 & 1.010 & 0.995 & 0.995 & 1.002 & 1.000 & 0.994 & 0.993 & 1.005 & 1.004 & 0.999 & 1.000 & 1.018 & 1.018 \\ \hline
mtams-mtams\_ & 0.945 & 0.973 & 0.764 & 0.620 & 0.971 & 0.990 & 0.956 & 0.562 & 0.968 & 0.986 & 0.979 & 0.602 & 0.903 & 0.969 & 0.763 & 0.542 \\ \hline
dtams-dtams\_ & 0.958 & 0.967 & 0.650 & 0.466 & 0.979 & 0.984 & 0.936 & 0.579 & 0.972 & 0.980 & 0.941 & 0.530 & 0.949 & 0.968 & 0.690 & 0.516 \\ \hline
mtsw-mtsw2 & 0.885 & 0.885 & 0.840 & 0.494 & 0.957 & 0.957 & 0.980 & 0.894 & 0.967 & 0.965 & 0.997 & 0.826 & 0.989 & 0.990 & 0.904 & 0.920 \\ \hline
mtsw\_-mtsw2\_ & 0.901 & 0.885 & 0.904 & 0.808 & 0.952 & 0.952 & 0.942 & 0.783 & 0.961 & 0.959 & 0.980 & 0.825 & 0.906 & 0.967 & 0.796 & 0.740 \\ \hline
mtsw2\_-dtsw2\_ & 0.820 & 0.746 & 0.871 & 0.970 & 0.848 & 0.784 & 0.836 & 0.932 & 0.841 & 0.779 & 0.914 & 0.938 & 0.930 & 0.855 & 0.943 & 0.968 \\ \hline
mtTS-mtsTS & 0.977 & 0.977 & 0.979 & 1.009 & 0.981 & 0.982 & 0.989 & 0.990 & 0.871 & 0.873 & 0.987 & 1.002 & 0.970 & 0.972 & 0.985 & 0.988 \\ \hline
dtTS-dtsTS & 0.979 & 0.980 & 0.976 & 1.010 & 0.956 & 0.957 & 0.956 & 0.990 & 0.889 & 0.905 & 0.926 & 0.992 & 0.962 & 0.949 & 1.010 & 1.019 \\ \hline
mtsTSkw-mtsTSpv & 0.962 & 0.962 & 0.967 & 0.953 & 0.968 & 0.969 & 0.980 & 0.966 & 0.961 & 0.961 & 0.965 & 0.998 & 0.974 & 0.976 & 0.984 & 0.981 \\ \hline
dtsTSkw-dtsTSpv & 0.969 & 0.967 & 0.981 & 1.003 & 0.973 & 0.975 & 0.959 & 0.925 & 0.948 & 0.945 & 0.976 & 0.955 & 0.956 & 0.966 & 0.952 & 0.973 \\ \hline
mtmT-mttmT & 0.962 & 0.957 & 0.996 & 1.001 & 0.991 & 0.991 & 0.997 & 1.002 & 0.877 & 0.872 & 0.995 & 1.000 & 0.995 & 0.996 & 1.016 & 1.023 \\ \hline
dtmT-dttmT & 0.989 & 0.976 & 0.997 & 1.015 & 0.982 & 0.980 & 0.984 & 0.994 & 0.874 & 0.863 & 0.960 & 0.996 & 0.992 & 0.997 & 1.007 & 0.994 \\ \hline
mlwss-dlwss & 0.804 & 0.809 & 0.792 & 0.648 & 0.852 & 0.852 & 0.889 & 0.823 & 0.841 & 0.837 & 0.925 & 0.854 & 0.904 & 0.909 & 0.802 & 0.906 \\ \hline
mtamH-mprof & 0.994 & 0.995 & 1.004 & 1.016 & 0.997 & 0.997 & 1.001 & 1.006 & 0.997 & 0.997 & 1.004 & 1.005 & 1.000 & 1.001 & 1.026 & 1.023 \\ \hline
dtamH-dprof & 0.996 & 0.997 & 1.001 & 1.016 & 0.999 & 0.999 & 1.003 & 1.000 & 0.999 & 0.999 & 1.006 & 1.003 & 1.002 & 1.003 & 1.021 & 1.022 \\ \hline
  \caption{Correlation of textual measures. See subsection~\ref{subsec:cor} for discussion and directions.}\label{tab:corTex}
\end{longtable*}

\newpage
\pagebreak
\newcolumntype{R}{>{}p{0.6cm}}
\clearpage
\begin{longtable*}{|p{2cm}|| c|c|c|c|| c|c|c|c|| c|c|c|c|| c|c|c|c|}\hline
  \centering
    \small
\setlength{\tabcolsep}{.06667em}
-\-  & \multicolumn{4}{c|}{CPP} & \multicolumn{4}{c|}{LAD} & \multicolumn{4}{c|}{LAU} & \multicolumn{4}{c|}{ELE} \\ \hline\hline
 & g. & p. & i. & h. &     g. & p. & i. & h. &    g. & p. & i. & h. &    g. & p. & i. & h. \\\hline \endfirsthead
-\-  & \multicolumn{4}{c|}{CPP} & \multicolumn{4}{c|}{LAD} & \multicolumn{4}{c|}{LAU} & \multicolumn{4}{c|}{ELE} \\ \hline\hline
 & g. & p. & i. & h. &     g. & p. & i. & h. &    g. & p. & i. & h. &    g. & p. & i. & h. \\\hline \endhead
ncont-$d_o$ & 0.923 & 0.335 & 0.459 & 0.934 & 0.836 & 0.405 & 0.265 & 0.744 & 0.851 & 0.455 & 0.476 & 0.775 & 0.510 & 0.682 & 0.371 & 0.414 \\ \hline
ncont-$s$ & 0.915 & 0.480 & 0.554 & 0.904 & 0.863 & 0.448 & 0.345 & 0.790 & 0.868 & 0.486 & 0.411 & 0.784 & 0.531 & 0.694 & 0.617 & 0.379 \\ \hline
ncont-$s_o$ & 0.907 & 0.420 & 0.568 & 0.888 & 0.858 & 0.462 & 0.380 & 0.781 & 0.874 & 0.484 & 0.515 & 0.793 & 0.575 & 0.776 & 0.826 & 0.457 \\ \hline
nc-$d$ & 0.930 & 0.362 & 0.259 & 0.930 & 0.921 & 0.316 & 0.368 & 0.877 & 0.921 & 0.371 & 0.209 & 0.866 & 0.592 & 0.604 & -0.064 & 0.380 \\ \hline
nc-$d_i$ & 0.923 & 0.220 & 0.151 & 0.917 & 0.907 & 0.180 & 0.259 & 0.852 & 0.900 & 0.209 & -0.113 & 0.831 & 0.535 & 0.266 & -0.302 & 0.285 \\ \hline
nc-$d_o$ & 0.929 & 0.377 & 0.342 & 0.942 & 0.918 & 0.346 & 0.400 & 0.877 & 0.922 & 0.396 & 0.480 & 0.876 & 0.616 & 0.733 & 0.398 & 0.463 \\ \hline
nc-$s$ & 0.951 & 0.441 & 0.359 & 0.962 & 0.932 & 0.353 & 0.415 & 0.905 & 0.923 & 0.400 & 0.310 & 0.878 & 0.734 & 0.695 & 0.517 & 0.620 \\ \hline
nc-$s_i$ & 0.946 & 0.258 & 0.207 & 0.961 & 0.915 & 0.208 & 0.291 & 0.881 & 0.896 & 0.224 & -0.100 & 0.842 & 0.717 & 0.296 & -0.311 & 0.600 \\ \hline
nc-$s_o$ & 0.951 & 0.458 & 0.448 & 0.957 & 0.938 & 0.378 & 0.427 & 0.916 & 0.936 & 0.431 & 0.545 & 0.895 & 0.704 & 0.805 & 0.834 & 0.568 \\ \hline
nc-$tri$ & 0.935 & 0.312 & 0.276 & 0.941 & 0.912 & 0.312 & 0.346 & 0.862 & 0.922 & 0.328 & 0.233 & 0.877 & 0.605 & 0.648 & 0.090 & 0.392 \\ \hline
nt-$d$ & 0.926 & 0.348 & 0.244 & 0.925 & 0.921 & 0.326 & 0.366 & 0.876 & 0.923 & 0.428 & 0.221 & 0.865 & 0.597 & 0.608 & -0.066 & 0.382 \\ \hline
nt-$d_i$ & 0.919 & 0.205 & 0.144 & 0.912 & 0.908 & 0.188 & 0.255 & 0.852 & 0.901 & 0.238 & -0.113 & 0.830 & 0.538 & 0.275 & -0.301 & 0.282 \\ \hline
nt-$d_o$ & 0.926 & 0.369 & 0.320 & 0.938 & 0.918 & 0.355 & 0.401 & 0.875 & 0.924 & 0.459 & 0.498 & 0.875 & 0.624 & 0.731 & 0.392 & 0.472 \\ \hline
nt-$s$ & 0.946 & 0.424 & 0.335 & 0.956 & 0.932 & 0.364 & 0.415 & 0.905 & 0.924 & 0.457 & 0.317 & 0.877 & 0.737 & 0.701 & 0.520 & 0.620 \\ \hline
nt-$s_i$ & 0.941 & 0.240 & 0.195 & 0.956 & 0.916 & 0.215 & 0.290 & 0.881 & 0.897 & 0.257 & -0.106 & 0.843 & 0.717 & 0.309 & -0.313 & 0.597 \\ \hline
nt-$s_o$ & 0.945 & 0.447 & 0.415 & 0.950 & 0.937 & 0.390 & 0.429 & 0.914 & 0.936 & 0.490 & 0.561 & 0.894 & 0.711 & 0.804 & 0.838 & 0.573 \\ \hline
nt-$bc$ & 0.865 & 0.247 & 0.085 & 0.845 & 0.851 & 0.128 & 0.180 & 0.774 & 0.857 & 0.173 & 0.086 & 0.768 & 0.501 & 0.305 & -0.185 & 0.238 \\ \hline
nt-$tri$ & 0.933 & 0.295 & 0.268 & 0.938 & 0.911 & 0.321 & 0.340 & 0.859 & 0.922 & 0.379 & 0.241 & 0.875 & 0.611 & 0.649 & 0.093 & 0.394 \\ \hline
ntd-$d$ & 0.905 & 0.430 & 0.402 & 0.903 & 0.917 & 0.437 & 0.428 & 0.860 & 0.921 & 0.557 & 0.256 & 0.863 & 0.827 & 0.708 & -0.039 & 0.409 \\ \hline
ntd-$d_i$ & 0.882 & 0.267 & 0.292 & 0.892 & 0.895 & 0.272 & 0.319 & 0.826 & 0.886 & 0.351 & -0.086 & 0.820 & 0.731 & 0.403 & -0.322 & 0.286 \\ \hline
ntd-$d_o$ & 0.925 & 0.443 & 0.468 & 0.912 & 0.924 & 0.458 & 0.446 & 0.871 & 0.939 & 0.563 & 0.521 & 0.885 & 0.882 & 0.780 & 0.486 & 0.536 \\ \hline
ntd-$s$ & 0.851 & 0.527 & 0.537 & 0.919 & 0.857 & 0.474 & 0.461 & 0.858 & 0.881 & 0.579 & 0.346 & 0.856 & 0.812 & 0.781 & 0.518 & 0.638 \\ \hline
ntd-$s_i$ & 0.833 & 0.322 & 0.372 & 0.914 & 0.842 & 0.298 & 0.340 & 0.837 & 0.847 & 0.365 & -0.084 & 0.818 & 0.735 & 0.436 & -0.346 & 0.574 \\ \hline
ntd-$s_o$ & 0.867 & 0.536 & 0.612 & 0.919 & 0.863 & 0.493 & 0.459 & 0.866 & 0.902 & 0.590 & 0.577 & 0.878 & 0.855 & 0.833 & 0.863 & 0.654 \\ \hline
ntd-$bc$ & 0.811 & 0.243 & 0.195 & 0.819 & 0.806 & 0.166 & 0.204 & 0.751 & 0.830 & 0.226 & 0.085 & 0.770 & 0.690 & 0.399 & -0.144 & 0.282 \\ \hline
ntd-$tri$ & 0.923 & 0.363 & 0.427 & 0.930 & 0.868 & 0.413 & 0.409 & 0.851 & 0.892 & 0.480 & 0.284 & 0.889 & 0.810 & 0.708 & 0.156 & 0.406 \\ \hline
ntd-$in\;cent$ & 0.523 & 0.036 & -0.019 & 0.451 & 0.631 & 0.096 & 0.105 & 0.318 & 0.666 & 0.123 & 0.103 & 0.367 & 0.583 & 0.138 & -0.007 & 0.158 \\ \hline
ntd-sector & 0.686 & 0.000 & 0.000 & 0.000 & 0.778 & 0.000 & 0.000 & 0.000 & 0.784 & 0.000 & 0.000 & 0.000 & 0.837 & 0.000 & 0.000 & 0.000 \\ \hline
ntd/nt-sector & -0.547 & 0.000 & 0.000 & 0.000 & -0.603 & 0.000 & 0.000 & 0.000 & -0.571 & 0.000 & 0.000 & 0.000 & -0.603 & 0.000 & 0.000 & 0.000 \\ \hline
mtsw2\_-sector & 0.555 & 0.000 & 0.000 & 0.000 & 0.546 & 0.000 & 0.000 & 0.000 & 0.502 & 0.000 & 0.000 & 0.000 & 0.683 & 0.000 & 0.000 & 0.000 \\ \hline
  \caption{Correlation of textual and topological measures. See subsection~\ref{subsec:cor} for discussion and directions.}\label{tab:corTexTop}
\end{longtable*}

\begin{table*}
  \centering
    \footnotesize
\setlength{\tabcolsep}{.26667em}
\begin{tabular}{|c|| c|c|c|c|| c|c|c|c|| c|c|c|c|| c|c|c|c|}\hline
-\-  & \multicolumn{4}{c|}{CPP} & \multicolumn{4}{c|}{LAD} & \multicolumn{4}{c|}{LAU} & \multicolumn{4}{c|}{ELE} \\ \hline\hline
 & g. & p. & i. & h. &     g. & p. & i. & h. &    g. & p. & i. & h. &    g. & p. & i. & h. \\\hline
$\lambda$ & 17.71 & 18.46 & 19.44 & 30.20 & 24.14 & 24.77 & 24.63 & 17.28 & 24.51 & 24.76 & 32.44 & 19.75 & 27.72 & 29.35 & 17.90 & 18.23 \\\hline
mtkwnsw\_ & 0.09 & -1.13 & -2.89 & 0.94 & -0.29 & -1.85 & 0.56 & 0.17 & -0.25 & 1.40 & -5.10 & 0.37 & 1.97 & -0.79 & 1.40 & -1.24 \\\hline
mtsw\_ & -0.34 & 1.56 & 0.86 & -0.09 & -0.37 & -2.48 & 1.97 & -0.55 & -1.80 & -2.90 & 1.88 & -2.21 & -0.85 & 1.08 & 6.66 & 3.57 \\\hline
mtsTS & 0.35 & -1.20 & -2.17 & 3.16 & -1.07 & -1.33 & 1.87 & 2.74 & 1.49 & -5.17 & -1.69 & 0.84 & -0.04 & -2.37 & -0.90 & 0.32 \\\hline
dtsTS & 0.45 & -1.20 & -1.49 & 0.63 & 0.34 & 1.72 & -0.51 & 1.06 & 0.47 & -2.67 & 1.50 & 5.08 & 1.47 & -1.28 & 1.12 & 0.32 \\\hline
mtsTSkw & -0.11 & -2.19 & 0.59 & -2.85 & 0.61 & 6.11 & -0.72 & -2.06 & 1.06 & -1.68 & -6.03 & -1.35 & 1.75 & 0.32 & -0.35 & 0.01 \\\hline
dtmT & 0.77 & 8.15 & -3.14 & 0.39 & 0.17 & -1.94 & -6.75 & 2.78 & 0.74 & 0.32 & 2.43 & -0.28 & -0.69 & -0.18 & 0.09 & 0.79 \\\hline
dttmT & -0.43 & 2.51 & 2.88 & 1.45 & -9.19 & 5.03 & -2.94 & -2.30 & 0.97 & -0.84 & -0.08 & -1.32 & -0.41 & -2.07 & 2.13 & 0.57 \\\hline
mtsmT & -0.45 & -2.04 & 1.25 & -1.22 & 6.29 & 6.63 & -1.57 & 0.03 & 2.87 & 1.75 & -2.80 & 0.37 & 1.54 & -8.42 & 2.13 & 0.57 \\\hline
dtsmT & -3.39 & 1.44 & -1.29 & 0.49 & 4.84 & -2.34 & -0.85 & 1.59 & -0.22 & -3.08 & -3.17 & -4.01 & -4.11 & -5.75 & -0.58 & -1.05 \\\hline
NN & -2.70 & 0.60 & -4.54 & -0.45 & -0.16 & 2.97 & 3.84 & -2.46 & 5.20 & -4.38 & -2.69 & -2.07 & 0.52 & 0.33 & -0.58 & -1.05 \\\hline
JJR & 0.54 & 3.13 & -0.51 & -0.51 & 1.51 & -0.94 & -1.40 & -2.82 & -0.00 & 6.45 & 0.72 & 1.21 & 1.78 & -1.02 & 0.13 & 0.10 \\\hline
JJS & -8.43 & 0.16 & -3.34 & 0.59 & 1.47 & 0.40 & -0.77 & -4.28 & -0.24 & -1.26 & 1.53 & -4.44 & 2.09 & 0.55 & 0.13 & 0.10 \\\hline
RB & 4.07 & 1.34 & 2.33 & 0.59 & 14.45 & -3.72 & -0.55 & 0.94 & -0.24 & 0.50 & -0.52 & 3.25 & 0.91 & -1.73 & 0.49 & -0.00 \\\hline
RBS & 0.49 & -1.54 & -0.87 & 0.66 & -3.89 & 0.99 & -0.82 & -0.26 & 7.90 & -0.53 & 0.54 & 1.11 & 7.59 & 0.81 & 0.27 & -0.92 \\\hline
VBD & 1.48 & 0.09 & -0.80 & -1.11 & -1.23 & 0.48 & 0.11 & -0.03 & 1.05 & -0.19 & 0.10 & 5.06 & 0.44 & -0.01 & 0.00 & 0.00 \\\hline
VBG & 0.19 & -0.68 & 1.10 & -1.29 & 0.18 & 0.89 & 0.37 & 0.90 & 5.63 & -0.40 & 0.46 & -4.17 & -1.61 & -1.30 & 0.00 & 0.00 \\\hline
IN & -1.74 & 0.32 & -0.56 & -1.24 & 1.55 & 0.44 & -0.23 & 1.27 & -5.74 & -0.44 & -1.66 & -0.60 & 1.09 & 0.32 & 0.00 & 0.00 \\\hline
PRP\$ & 6.51 & 0.22 & 0.29 & 0.41 & -0.77 & -0.18 & -0.11 & 0.25 & -0.26 & 0.22 & 0.14 & -0.08 & 0.14 & 0.34 & 0.00 & 0.00 \\\hline
PDT & 5.12 & 0.11 & 0.24 & 0.41 & -0.14 & -0.15 & 0.19 & -0.25 & 0.31 & 0.56 & -0.43 & 0.40 & -1.40 & -0.37 & 0.00 & 0.00 \\\hline
\end{tabular}
  \caption{Composition of first component (threshold: $|val|>0.05$). See subsection~\ref{subsec:pc} for discussion and directions.}\label{tab:pca1}
\end{table*}

\begin{table*}
    \centering
    \footnotesize
\setlength{\tabcolsep}{.26667em}
\begin{tabular}{|c|| c|c|c|c|| c|c|c|c|| c|c|c|c|| c|c|c|c|}\hline
-\-  & \multicolumn{4}{c|}{CPP} & \multicolumn{4}{c|}{LAD} & \multicolumn{4}{c|}{LAU} & \multicolumn{4}{c|}{ELE} \\ \hline\hline
 & g. & p. & i. & h. &     g. & p. & i. & h. &    g. & p. & i. & h. &    g. & p. & i. & h. \\\hline
$\lambda$ & 11.48 & 10.25 & 12.23 & 15.72 & 11.97 & 9.39 & 12.05 & 13.19 & 11.87 & 9.76 & 11.42 & 11.26 & 12.49 & 11.17 & 15.75 & 13.31 \\\hline
nc & -5.21 & -3.60 & 3.06 & -4.12 & 4.62 & -3.49 & 2.30 & 0.94 & 4.49 & 3.76 & -2.75 & -0.53 & 2.87 & -2.66 & -3.97 & -0.03 \\\hline
Nwsssw/Nwss & 0.43 & 0.34 & 0.35 & -0.66 & 0.21 & -0.10 & 0.93 & -0.30 & 0.07 & 0.20 & 0.23 & -0.54 & 1.10 & -0.44 & -5.11 & 4.46 \\\hline
dtsmT & -5.57 & 0.29 & 1.79 & -1.41 & 2.42 & 3.05 & 2.04 & -0.28 & -0.33 & 0.01 & -0.16 & -1.67 & 3.51 & 1.08 & 0.50 & 1.02 \\\hline
JJR & 0.22 & -4.19 & 0.73 & 1.98 & 0.14 & 1.31 & -0.09 & -0.11 & -0.53 & -5.38 & -1.45 & 1.64 & -0.69 & 0.81 & 0.63 & 0.11 \\\hline
RB & -0.68 & -2.74 & 0.90 & 0.22 & -6.12 & 2.64 & 0.65 & -0.29 & -0.16 & -0.89 & -1.69 & -1.06 & 0.30 & -0.63 & 0.21 & -0.52 \\\hline
IN & 0.46 & 0.98 & -0.35 & -0.54 & -2.09 & 1.97 & 0.19 & -0.68 & 5.08 & 2.39 & 2.17 & -0.31 & -0.45 & -1.22 & 0.00 & 0.00 \\\hline
WP\$ & -0.15 & 1.52 & 2.10 & 0.00 & -5.22 & -0.63 & -2.09 & 0.13 & 4.57 & -1.19 & -1.36 & -0.94 & 0.21 & 0.12 & -0.05 & -1.11 \\\hline
CD & -5.85 & 0.15 & 1.02 & 0.00 & 0.78 & 0.70 & -4.05 & -0.25 & 0.05 & 0.88 & -3.35 & -1.58 & -0.48 & -0.76 & 0.03 & -1.11 \\\hline
mtamH & 0.31 & -0.93 & -3.67 & 0.00 & -0.57 & 3.99 & 1.73 & 1.03 & -0.23 & -0.76 & -2.07 & 6.02 & -0.14 & 2.64 & 0.41 & 0.05 \\\hline
dtamH & 0.10 & -0.35 & -1.14 & 0.21 & 0.51 & -1.88 & -5.96 & -0.50 & -0.31 & -6.22 & 1.31 & 1.95 & -0.74 & -0.98 & 0.41 & 0.05 \\\hline
mprof & -3.63 & 2.72 & -1.61 & 0.21 & 0.20 & 0.72 & -0.57 & -2.39 & 0.81 & -1.40 & 0.49 & 0.56 & 0.26 & 5.42 & -0.06 & -0.29 \\\hline
dprof & -0.73 & 1.16 & -2.33 & 0.24 & -0.52 & 0.75 & 0.44 & -1.12 & 1.18 & 3.76 & 7.77 & 2.01 & -0.19 & -5.61 & -0.06 & -0.29 \\\hline
$d_o$ & -0.02 & -0.01 & -4.02 & 1.06 & 0.29 & 0.64 & -0.65 & 8.56 & 0.39 & 1.44 & 0.35 & -1.46 & -0.12 & 1.40 & 0.03 & -0.28 \\\hline
$s_o$ & -0.71 & 2.39 & -1.52 & 0.25 & 0.20 & 8.51 & -0.11 & 0.23 & -1.57 & -6.21 & 1.25 & -0.36 & -1.08 & 0.47 & -0.04 & 0.08 \\\hline
$bc$ & 0.70 & 1.55 & 0.11 & -0.18 & -11.23 & -0.12 & -0.17 & 0.05 & -11.59 & 0.59 & 0.42 & 0.76 & -9.01 & -0.59 & -0.20 & 0.78 \\\hline
$tri$ & -0.07 & 8.58 & 0.00 & -0.45 & -5.88 & -0.46 & 0.00 & 0.00 & -2.17 & 0.02 & 0.00 & -0.00 & -3.52 & -0.25 & -0.98 & 0.03 \\\hline
$in\;cent$ & 15.09 & -0.00 & 0.00 & 0.53 & -0.06 & 0.00 & 0.00 & 0.00 & 0.15 & -0.00 & 0.00 & 0.00 & -0.79 & -0.00 & 0.06 & 1.25 \\\hline
\end{tabular}
  \caption{Composition of second component (threshold: $|val|>0.05$). See subsection~\ref{subsec:pc} for discussion and directions.}\label{tab:pca2}
\end{table*}

\begin{table*}
\begin{tabular}{|c|| c|c|c|c|| c|c|c|c|| c|c|c|c|| c|c|c|c|}\hline
  \centering
    \scriptsize
\setlength{\tabcolsep}{.26667em}
-\-  & \multicolumn{4}{c|}{CPP} & \multicolumn{4}{c|}{LAD} & \multicolumn{4}{c|}{LAU} & \multicolumn{4}{c|}{ELE} \\ \hline\hline
 & g. & p. & i. & h. &     g. & p. & i. & h. &    g. & p. & i. & h. &    g. & p. & i. & h. \\\hline
$\lambda$ & 8.97 & 7.53 & 7.71 & 7.19 & 8.28 & 8.07 & 7.97 & 10.58 & 8.45 & 7.40 & 6.61 & 8.45 & 6.56 & 6.01 & 11.01 & 9.85 \\\hline
Nkw/nt & 1.81 & 1.26 & -1.38 & 0.76 & -0.58 & -0.52 & -0.32 & -3.03 & 0.85 & 1.13 & -1.00 & -5.90 & -1.77 & -1.15 & 5.75 & -0.20 \\\hline
mtsw2\_ & 2.46 & 1.66 & -1.23 & 3.90 & 0.17 & 0.55 & -1.23 & 0.16 & 0.60 & -1.59 & 0.10 & 6.54 & 0.27 & 0.86 & 0.19 & -1.89 \\\hline
mtsTS & 0.42 & -0.97 & -1.76 & 2.08 & -6.73 & 1.74 & -1.53 & -1.63 & 0.74 & -2.66 & -1.48 & 0.68 & 2.08 & -4.13 & 0.35 & 0.51 \\\hline
dtsTSkw & 1.66 & 2.31 & 1.25 & 0.93 & 3.90 & -5.60 & -3.81 & 0.01 & 3.03 & 2.75 & -1.19 & -0.69 & 0.71 & 1.29 & 1.01 & 0.72 \\\hline
mtsTSpv & 0.71 & 6.83 & 1.68 & -0.24 & 2.40 & -0.32 & -0.64 & 0.14 & 3.85 & -2.97 & 1.10 & 3.45 & -2.74 & 0.70 & -1.28 & -0.08 \\\hline
dtsTSpv & 5.50 & -2.29 & -2.81 & 0.93 & -3.36 & 3.78 & 0.86 & -0.23 & -2.61 & 1.40 & -2.96 & 1.31 & -1.59 & -3.64 & -1.06 & -1.12 \\\hline
mtmT & -2.90 & 0.60 & -0.74 & 1.75 & 5.53 & -2.22 & -2.11 & 0.37 & -2.22 & -2.90 & 2.63 & -2.68 & -0.44 & 2.35 & -1.06 & -1.12 \\\hline
dtmT & 1.64 & -0.43 & 0.52 & 0.21 & 0.56 & -5.70 & 0.33 & 2.14 & 7.11 & -3.32 & -1.85 & -2.08 & -2.93 & 3.07 & 0.43 & -0.43 \\\hline
dttmT & 0.90 & -0.20 & -2.54 & 2.69 & 3.68 & 3.77 & -0.22 & -0.18 & -3.64 & 2.06 & 1.57 & 0.67 & 5.92 & 8.07 & -1.52 & -1.03 \\\hline
mtsmT & -0.33 & -5.56 & 4.20 & 1.39 & 5.68 & -4.48 & 2.97 & 2.34 & 3.76 & -7.62 & 0.84 & 1.13 & 9.98 & 2.51 & -1.52 & -1.03 \\\hline
dtsmT & -1.96 & 0.77 & 2.26 & -0.26 & 1.79 & 0.92 & 0.06 & -0.03 & 5.00 & 1.80 & -2.45 & 2.24 & 1.99 & -5.87 & 0.34 & 1.25 \\\hline
NN & 2.78 & 0.13 & 0.15 & 1.10 & -0.60 & 6.82 & -0.07 & -0.10 & -2.70 & 4.95 & 1.05 & -1.69 & 0.46 & 3.08 & 0.34 & 1.25 \\\hline
\end{tabular}
  \caption{Composition of third component (threshold: $|val|>0.05$). See subsection~\ref{subsec:pc} for discussion and directions.}\label{tab:pca3}
\end{table*}

\begin{table*}
\begin{tabular}{|c|| c|c|c|c|| c|c|c|c|| c|c|c|c|| c|c|c|c|}\hline
  \centering
    \scriptsize
\setlength{\tabcolsep}{.06667em}
-\-  & \multicolumn{4}{c|}{CPP} & \multicolumn{4}{c|}{LAD} & \multicolumn{4}{c|}{LAU} & \multicolumn{4}{c|}{ELE} \\ \hline\hline
 & g. & p. & i. & h. &     g. & p. & i. & h. &    g. & p. & i. & h. &    g. & p. & i. & h. \\\hline
$\lambda$ & 3.81 & 4.00 & 6.75 & 5.52 & 4.26 & 4.03 & 5.79 & 6.03 & 3.55 & 3.68 & 5.29 & 7.14 & 5.65 & 5.66 & 7.78 & 8.31 \\\hline
ncont & 2.30 & 2.60 & -2.75 & -2.62 & -3.27 & -3.23 & -3.16 & -1.59 & -3.55 & 3.47 & -2.74 & 1.75 & -2.87 & -3.16 & 5.14 & 1.67 \\\hline
dtsw & 0.58 & -0.12 & -0.84 & -1.52 & 0.38 & -0.19 & 0.27 & -0.83 & -0.06 & 0.10 & 0.99 & -0.62 & -0.04 & -0.34 & 12.71 & 1.12 \\\hline
WRB & -1.39 & 1.62 & -5.39 & 0.00 & -5.16 & 4.23 & 0.15 & 0.36 & 0.52 & -1.58 & 1.69 & 1.24 & 1.54 & -5.08 & 0.07 & 0.00 \\\hline
WP & -2.35 & 1.63 & -1.44 & 0.00 & 3.75 & 1.90 & -2.32 & -0.36 & 7.41 & -5.00 & -1.92 & -0.65 & -2.44 & -0.32 & -0.62 & -0.90 \\\hline
WP\$ & -4.35 & 2.81 & -4.42 & 0.00 & 2.29 & -0.54 & -1.24 & -0.70 & -3.39 & -5.06 & -0.68 & 0.61 & 0.54 & -2.12 & -0.62 & 1.60 \\\hline
EX & -7.04 & 8.54 & -0.88 & 0.00 & -0.70 & 1.26 & -1.12 & -3.15 & 0.36 & -0.35 & -0.98 & -2.34 & -0.44 & -0.51 & -0.23 & -0.69 \\\hline
UH & 0.99 & -1.17 & -7.02 & 0.00 & 0.55 & -0.90 & -0.68 & 0.27 & -0.06 & -0.13 & 1.79 & 0.37 & -0.19 & -0.29 & 0.44 & -0.69 \\\hline
FW & -1.35 & 5.80 & 1.68 & 0.00 & -0.24 & 0.45 & 1.10 & -2.52 & -0.21 & 1.46 & -0.64 & -1.17 & -0.06 & -0.62 & 0.44 & 1.35 \\\hline
mlwss & 0.53 & 1.41 & -0.79 & 0.00 & 1.00 & -0.55 & 0.48 & 3.77 & -0.93 & -0.42 & -1.96 & 6.52 & -0.39 & 0.34 & 0.02 & 1.66 \\\hline
dlwss & 2.83 & 0.44 & -1.10 & 0.00 & 0.01 & 0.47 & 2.94 & -0.95 & -1.30 & 0.85 & 2.89 & 6.23 & 0.82 & 1.81 & 0.42 & 1.66 \\\hline
dprof & -1.24 & 1.04 & -0.39 & -0.81 & -0.38 & 0.56 & -0.08 & 5.71 & -0.20 & -0.17 & -1.31 & -0.52 & -0.37 & -0.03 & 0.19 & 0.34 \\\hline
$d$ & 0.43 & -0.91 & -1.59 & -0.81 & -0.21 & 0.42 & -2.60 & -6.63 & 0.49 & 0.38 & 4.88 & 0.20 & 1.43 & 1.49 & 0.65 & 0.58 \\\hline
$s$ & 0.07 & -0.50 & -0.46 & 0.80 & 0.62 & -1.12 & 2.74 & 3.75 & 0.30 & -0.07 & -6.44 & 1.16 & 1.27 & -1.08 & -0.87 & 0.23 \\\hline
$s_i$ & -0.34 & 0.18 & -0.54 & 1.27 & -1.05 & -0.90 & -13.22 & 1.39 & 0.40 & -1.10 & 9.86 & 1.39 & 0.62 & -1.29 & -0.86 & 0.23 \\\hline
$bc$ & -0.17 & -0.06 & 0.03 & -0.71 & -0.11 & -0.14 & -1.03 & -0.92 & 0.02 & 0.28 & 5.11 & -0.56 & -2.46 & 1.63 & -0.34 & 0.46 \\\hline
$tri$ & 1.09 & 0.11 & 0.00 & -0.31 & 0.11 & 14.18 & -0.00 & -0.00 & 1.23 & -0.40 & -0.00 & 0.00 & 3.12 & -1.50 & -0.49 & -0.24 \\\hline
$cv$ & -0.24 & 0.24 & 0.00 & -0.25 & -2.62 & -4.41 & 0.00 & 0.00 & 0.34 & -15.74 & 0.00 & 0.00 & 14.09 & 16.09 & -0.36 & -0.24 \\\hline
$in\;cent$ & 0.02 & -0.00 & 0.00 & -0.25 & 14.87 & 0.00 & 0.00 & 0.00 & 16.34 & -0.00 & -0.00 & 0.00 & 3.57 & -0.00 & -0.36 & -0.51 \\\hline
\end{tabular}
  \caption{Composition of fourth component (threshold: $|val|>0.05$). See subsection~\ref{subsec:pc} for discussion and directions.}\label{tab:pca4}
\end{table*}

\begin{table*}
\begin{tabular}{|c|| c|c|c|c|| c|c|c|c|| c|c|c|c|| c|c|c|c|}\hline
  \centering
    \scriptsize
\setlength{\tabcolsep}{.06667em}
-\-  & \multicolumn{4}{c|}{CPP} & \multicolumn{4}{c|}{LAD} & \multicolumn{4}{c|}{LAU} & \multicolumn{4}{c|}{ELE} \\ \hline\hline
 & g. & p. & i. & h. &     g. & p. & i. & h. &    g. & p. & i. & h. &    g. & p. & i. & h. \\\hline
$\lambda$ & 3.48 & 3.38 & 4.54 & 4.16 & 3.42 & 3.74 & 4.60 & 4.41 & 2.97 & 3.14 & 3.83 & 5.33 & 3.32 & 3.58 & 5.95 & 7.29 \\\hline
Nkwnssnsw/Nkw & -6.10 & -3.15 & 0.31 & 0.94 & -2.48 & -1.57 & -0.20 & 0.43 & 1.12 & 1.63 & 0.33 & 1.23 & -2.41 & -0.97 & -1.30 & 1.91 \\\hline
mtsTS & -4.04 & -1.67 & -1.79 & 1.54 & -1.90 & 0.84 & -5.02 & -0.15 & -1.53 & -1.19 & 0.50 & 0.67 & 1.56 & -0.20 & 0.89 & 0.24 \\\hline
dtsTS & -0.97 & 3.59 & -1.47 & 0.09 & 5.70 & 0.45 & 2.78 & 0.45 & -3.19 & 6.97 & 1.40 & 0.12 & -1.89 & -1.59 & -0.20 & 0.30 \\\hline
dtsTSkw & 0.03 & -1.95 & 0.87 & -1.14 & -5.87 & -2.23 & 0.06 & -2.74 & -6.87 & -4.49 & -0.07 & 1.36 & 2.53 & -0.85 & -0.44 & -0.20 \\\hline
dtsTSpv & 1.87 & -0.41 & 0.82 & 3.16 & 2.28 & -5.32 & -1.16 & 1.35 & 4.05 & -4.68 & 3.13 & -0.67 & 3.68 & 3.04 & -0.58 & -0.84 \\\hline
sector & 0.00 & 0.00 & 0.00 & 0.48 & 0.00 & 0.00 & 0.00 & -0.00 & 0.00 & 0.00 & 0.00 & 0.00 & -0.00 & -0.00 & 11.05 & 1.52 \\\hline
\end{tabular}
  \caption{Composition of fifth component (threshold: $|val|>0.05$). See subsection~\ref{subsec:pc} for discussion and directions.}\label{tab:pca5}
\end{table*}

\clearpage
\newpage
\section{Histograms of existent and incident words}\label{sec:resE}
\begin{figure*}[h!]
    \centering
    \includegraphics[width=\textwidth]{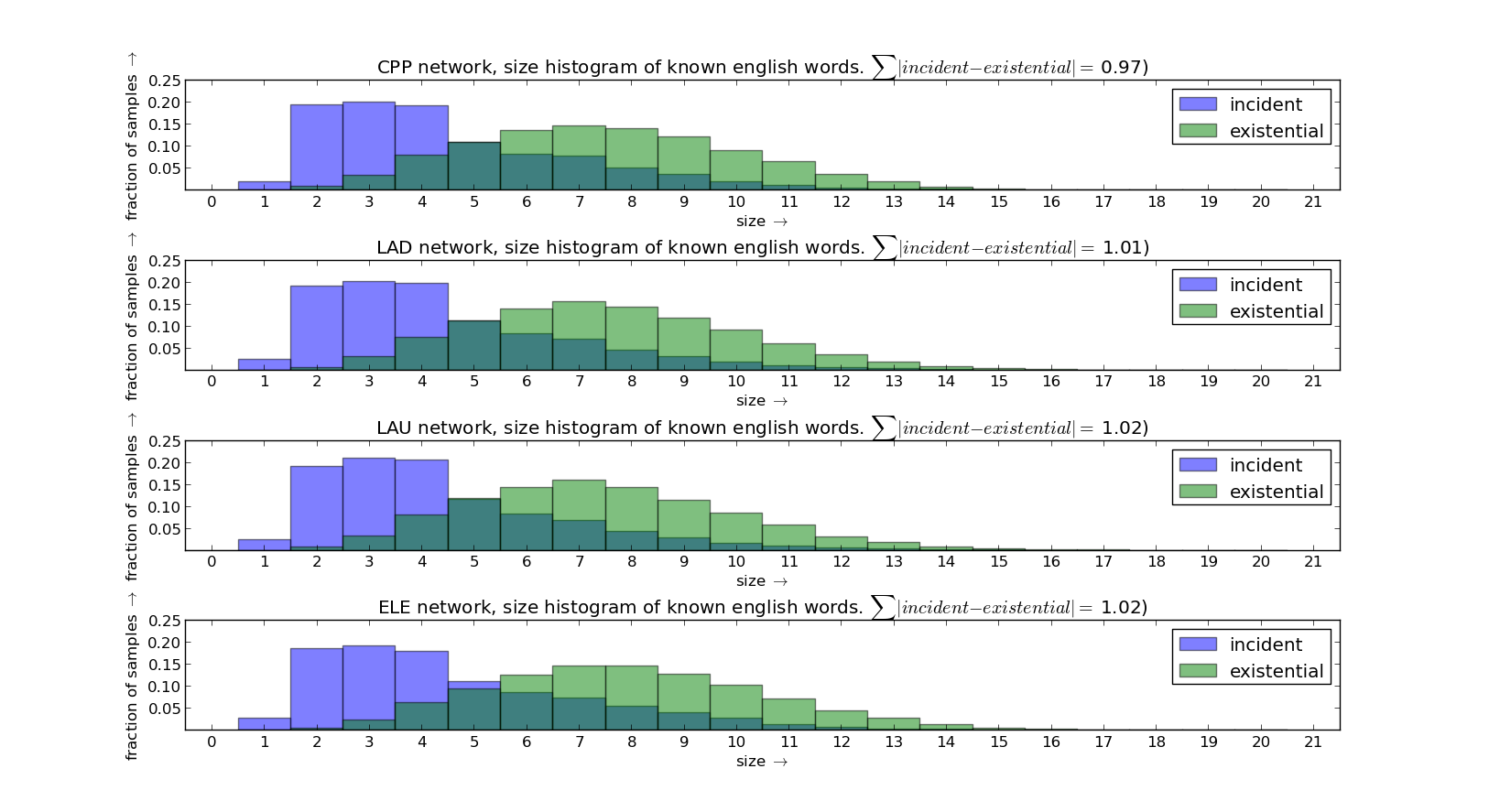}
    \caption{Size of words that are known in English. Crossing of incident and existential sizes is around 5 (Figure~\ref{fig:kwnsw} shows a shift to length 6-7 when consider only non stopwords). Words with three letters have maximum incidence, while most words have 7 letters. See subsection~\ref{subsec:sii} for discussion and directions.}
    \label{fig:kw}
\end{figure*}

\begin{figure*}[h!]
    \centering
    \includegraphics[width=\textwidth]{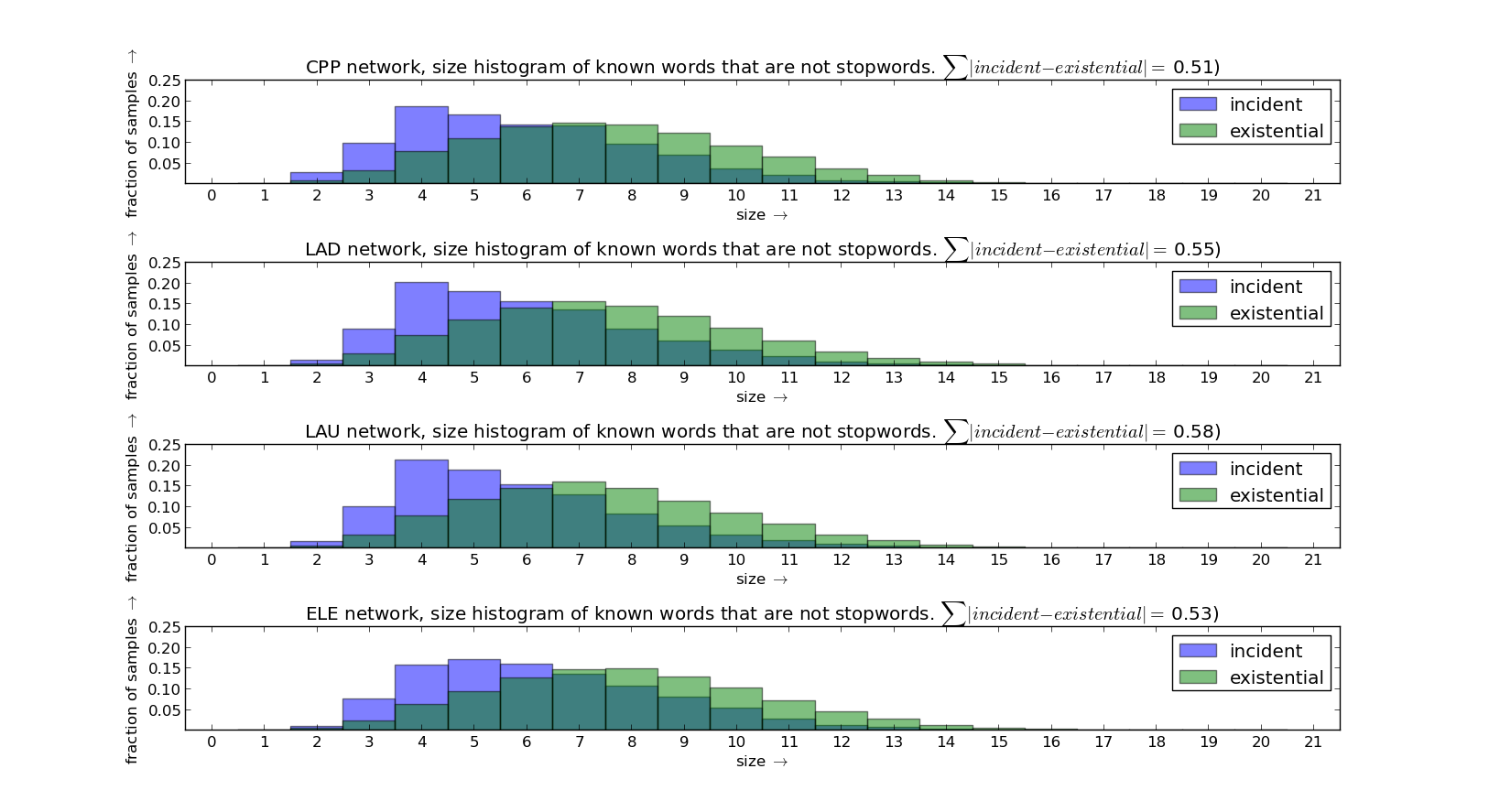}
    \caption{Size of words that are known in English and are not stopwords. Crossing of incident and existential sizes is around 6-7 (figure~\ref{fig:kw} shows a shift to length 5 when considered stopwords). In this case, words with 4 letters have maximum incidence, while most words still have 7 letters. Exception for ELE, which exhibits maximum incidence of words with 5 letters and most words having 8 letters, which might be associated with ELE network typology discussed in tables~\ref{tab:tokens} and~\label{tab:caracteres}. See subsection~\ref{subsec:sii} for discussion and directions.}
    \label{fig:kwnsw}
\end{figure*}

\begin{figure*}[h!]
    \centering
    \includegraphics[width=\textwidth]{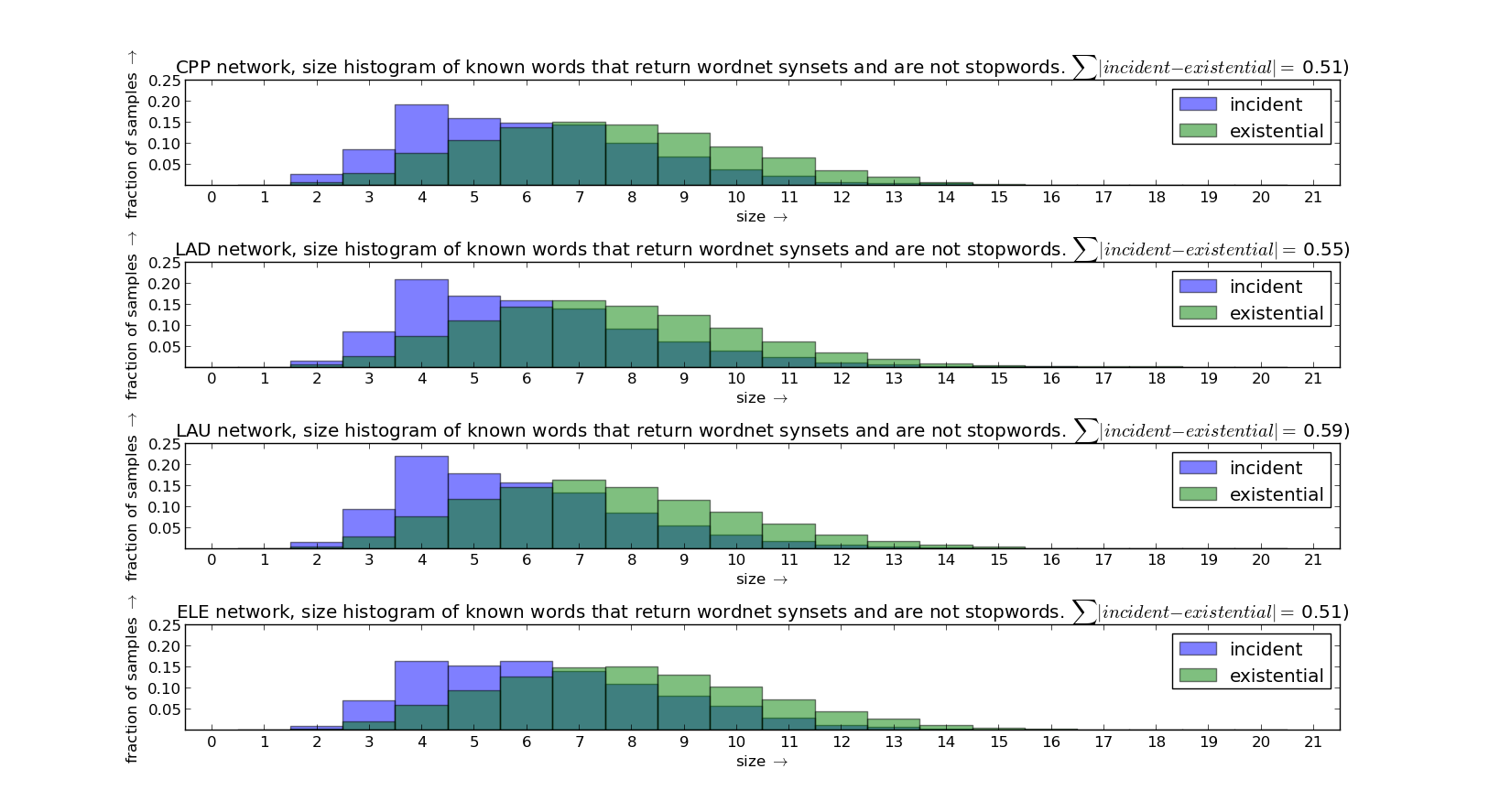}
    \caption{Size of words that are known, are not stopwords and have synsets. Resembles figure~\ref{fig:kwnsw}. Stopword sizes histogram are in figure~\ref{fig:sw}. Differences suggests $\approx 0.5$ might be constant. LAD and LAU exquisite vocabulary (GNU/Linux, programming, sound/signal processing, music) might be responsible for higher difference of distributions. See subsection~\ref{subsec:sii} for discussion and directions. See subsection~\ref{subsec:sii} for discussion and directions.}
    \label{fig:kwssnsw}
\end{figure*}

\begin{figure*}[h!]
    \centering
    \includegraphics[width=\textwidth]{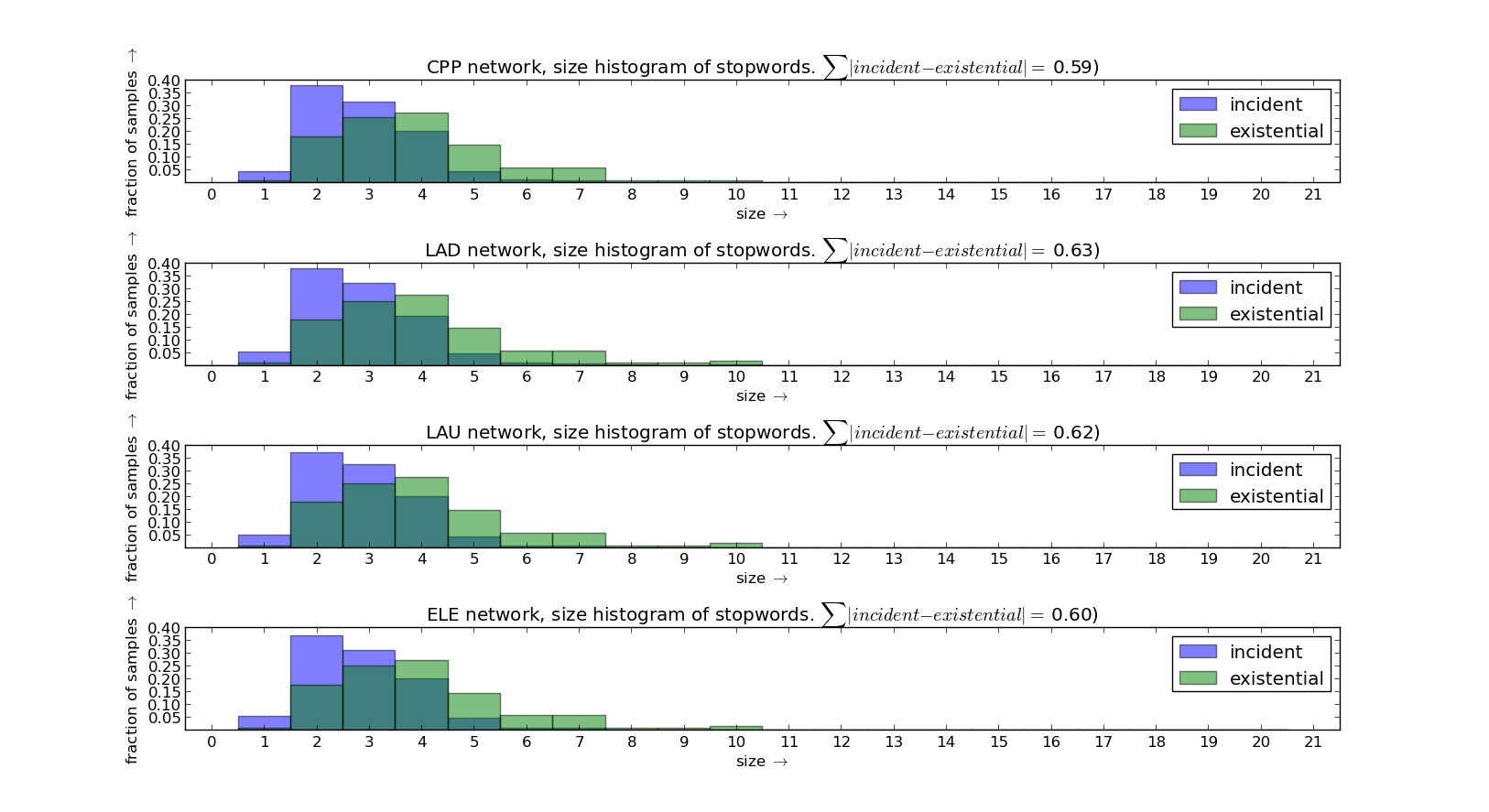}
    \caption{Size histogram of stopwords. Stopwords with two letters are the most frequent, while most of them have four letters. Differences in distribution seem stable around $\approx 0.6$. See subsection~\ref{subsec:sii} for discussion and directions.}
    \label{fig:sw}
\end{figure*}

\begin{figure*}[h!]
    \centering
    \includegraphics[width=\textwidth]{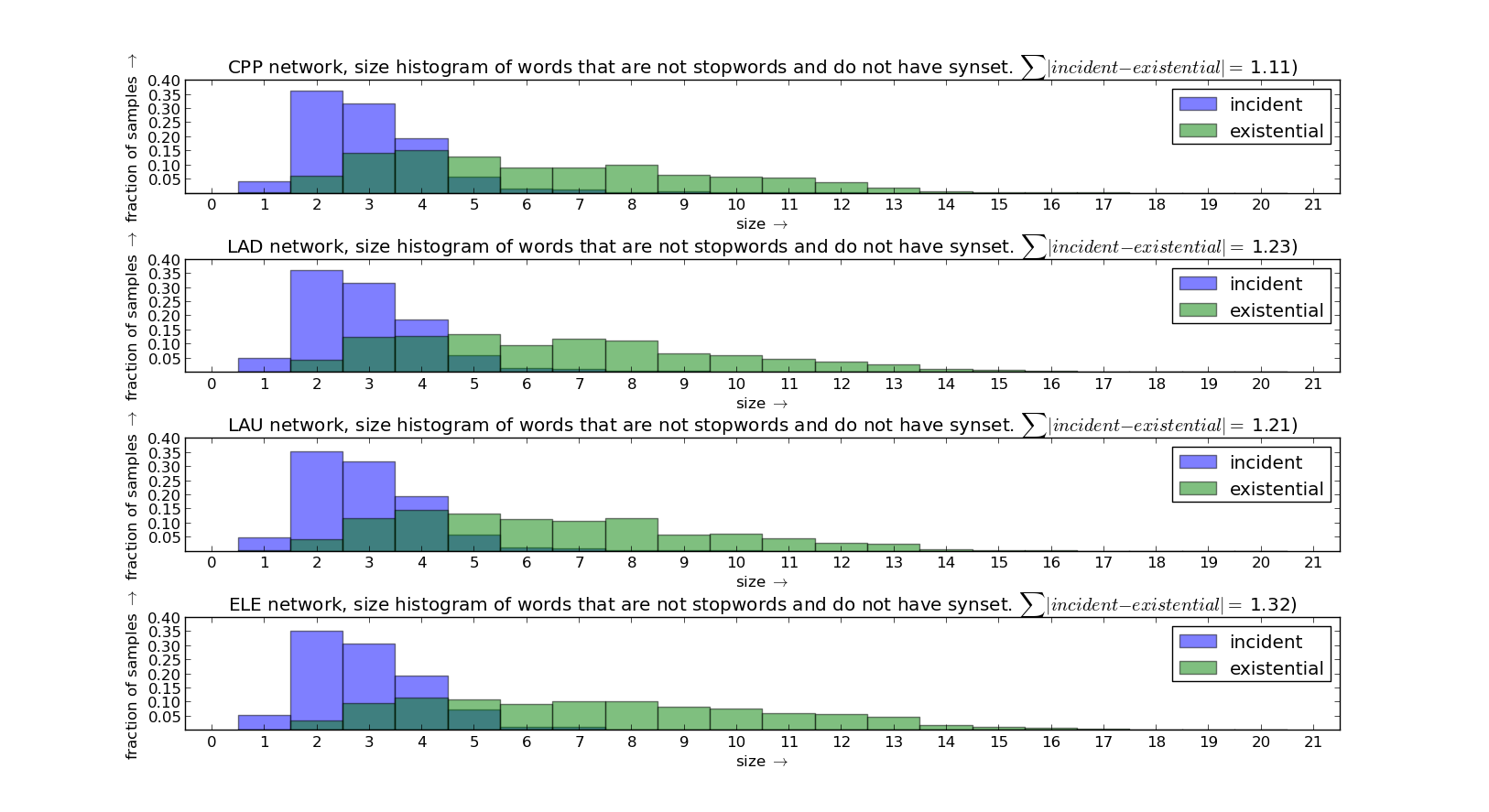}
    \caption{Size histogram of known English words that are not stopwords and do not return synsets. Differences in distribution suggests less stable behavior, with high incidence of few words high number of existing words with many letters. Observe difference $\geq 1$, as observed only with all known words, but even higher. See subsection~\ref{subsec:sii} for discussion and directions.}
    \label{fig:nssnsw}
\end{figure*}

 See subsection~\ref{subsec:sii}, and Figures~\ref{fig:kw}-\ref{fig:nssnsw} for discussion and directions.

\clearpage
\section{Online scripts, data and writing}\label{sec:on}
All data can be accessed in the GMANE database, which consists of some tenths of thousands of mailing lists, with constant updates~\cite{GMANE}. All scripts used on this article are in \url{http://sourceforge.net/p/labmacambira/fimDoMundo/ci/master/tree/python/toolkitGMANE/}. The git repository of this article itself is \url{https://github.com/ttm/artigoTextoNasRedes.git}.

\bibliography{paper}

\begin{thebibliography}{10}

\bibitem{Note1}
gmane.comp.gcc.libstdc++.devel is list ID in GMANE archive.

\bibitem{Note2}
gmane.linux.audio.devel is list ID in GMANE archive.

\bibitem{Note3}
gmane.linux.audio.users is list ID in GMANE archive.

\bibitem{Note4}
gmane.politics.election-methods is list ID in GMANE archive.

\bibitem{OPS}
Ontologia de participação social.
\newblock \url{http://tinyurl.com/p2doueu}.

\bibitem{pnud5}
{\em Produto 5 da consultoria PNUD/ONU de Renato Fabbri}.
\newblock
  \url{https://github.com/ttm/pnud4/blob/master/latex/produto.pdf?raw=true}.

\bibitem{Deborah}
D.~C. Antunes, A.~A. Zuin, et~al.
\newblock Do bullying ao preconceito: os desafios da barb{\'a}rie {\`a}
  educa{\c{c}}{\~a}o.
\newblock {\em Psicologia \& Sociedade}, 20(1):33--42, 2008.

\bibitem{nltkBook}
S.~Bird, E.~Klein, and E.~Loper.
\newblock {\em Natural language processing with Python}.
\newblock " O'Reilly Media, Inc.", 2009.

\bibitem{pbr}
R.~P. Cavalcante.
\newblock Influ{\^e}ncia dos movimentos sociais na forma{\c{c}}{\~a}o da agenda
  pol{\'\i}tica: as jornadas de junho de 2013 no brasil.
\newblock 2014.

\bibitem{easley}
D.~Easley and J.~Kleinberg.
\newblock {\em Networks, crowds, and markets: Reasoning about a highly
  connected world}.
\newblock Cambridge University Press, 2010.

\bibitem{evoSN}
R.~F. et~al.
\newblock Stability in human interaction networks: primitive typology of
  vertex, prominance of measures and activity statistics, May 2014.
\newblock
  \url{http://sourceforge.net/p/labmacambira/fimDoMundo/ci/master/tree/textos/%
evolutionSN/paper.pdf?format=raw}.

\bibitem{ensaio}
R.~Fabbri.
\newblock Ensaio sobre o auto-aproveitamento: um relato de investidas naturais
  na participação social.
\newblock \url{http://arxiv.org/abs/1412.6868}.

\bibitem{letrasMachado}
R.~Fabbri.
\newblock Incidência de letras na obra de machado de assis, 2012.
\newblock
  \url{http://sourceforge.net/p/labmacambira/rcpln/ci/master/tree/pln/trabLetr%
as/resumoLetras.pdf?format=raw}.

\bibitem{GMANE}
K.~Marek-Spartz, P.~Chesley, and H.~Sande.
\newblock Construction of the gmane corpus for examining the diffusion of
  lexical innovations.
\newblock 2012.

\end{thebibliography}

\end{document}